\documentclass[11pt]{article}

\usepackage[margin=1in]{geometry}
\usepackage{setspace}
\usepackage{amsmath, amssymb}
\usepackage{graphicx}
\usepackage{booktabs}
\usepackage{natbib}
\usepackage{hyperref}

\usepackage{amsthm}

\usepackage{xcolor}

\theoremstyle{plain}
\newtheorem{proposition}{Proposition}

\theoremstyle{remark}

\hypersetup{colorlinks=true, linkcolor=blue, citecolor=blue, urlcolor=blue}
\IfFileExists{placeins.sty}{\usepackage{placeins}}{\providecommand{\FloatBarrier}{}}

\title{\textbf{Demand Estimation with Variable Choice Sets: \\ 
A Likelihood Correction for Nested Logit} \thanks{We thank Partha Deb and Jessica Van Parys for helpful input. A Stata package implementing the estimator, \textit{exactnl}, accompanies the paper. All errors are our own.} }

\author{
Matthew J. Baker \\
\small Department of Economics, Hunter College and the Graduate Center, CUNY \\
\small \texttt{matthew.baker@hunter.cuny.edu}
\and
Lisa M. George \\
\small Department of Economics, Hunter College and the Graduate Center, CUNY \\
\small \texttt{lisa.george@hunter.cuny.edu}
}

\date{August 21, 2026}

\begin{document}

\maketitle

\begin{abstract}
We derive the exact likelihood of the Berry (1994) share-form nested logit and show it contains a Jacobian term that depends on nest size and the nesting parameter. Omitting the term biases within-nest substitution estimates wherever choice sets vary across markets. Commonly-used count instruments for the within-nest share fail exclusion when product counts enter demand directly. The corrected likelihood identifies substitution without them, making crowding estimable. We apply the estimator to Medicare Advantage plan proliferation after 2019 elimination of the ``meaningful difference'' requirement. Estimates indicate  plan proliferation raised consumer surplus in most markets, but ignoring crowding overstates the gains.

\vspace{1em}


\vspace{1em}


\end{abstract}

\vspace{1em}
\newpage 
\section{Introduction}

The nested multinomial logit model is a workhorse of empirical demand analysis
in differentiated product markets. Its appeal rests on the fact that when applied
to aggregate market shares, the model admits a linear estimating equation.
Specifically, \citet{berry94} showed that the log ratio of product $j$'s market
share to the outside good share can be written as
\begin{equation} \label{berry_inversion}
\ln s_j - \ln s_0 = \mu_{jg} + \sigma_g \ln s_{j|g} + \xi_{j},
\end{equation}
where $\mu_{jg}$ is the systematic part of mean utility, $\sigma_g \in [0,1)$ is
the nesting parameter governing within-group substitution in nest $g$, $s_{j|g}$ is product
$j$'s share within group $g$, and $\xi_{j}$ is unobserved product quality. This share
inversion, extended by \citet{blp95} to form what is conventionally referred to
as the Berry-Levinsohn-Pakes (BLP) approach, forms the basis of virtually all
subsequent work on demand for differentiated products.

In this paper, we derive the exact likelihood of the Berry inversion for the nested logit model. We show that the likelihood contains a Jacobian term that depends on the number of products in each nest and the nesting parameter. The term is absent from standard nested-logit share-equation estimators. Because it depends on the nesting parameter, omission of the Jacobian biases likelihood estimation even when choice sets are fixed. When choice sets vary, the term scales with nest size, so the resulting bias varies systematically with the same product-count variation commonly used to instrument the within-group share. This connection reveals a second problem with standard practice: count instruments fail exclusion whenever product counts enter demand directly or proxy for omitted conditions such as crowding in product space.    In this sense, our result makes concrete the concern raised by \citet{ackerberg05}, that logit, nested logit, and BLP models impose restrictions on how unobserved product space changes as the number of products changes.  We illustrate the approach with an application to Medicare Advantage plan choice, using county-year enrollment data around a 2019 policy change that substantially increased the number of plans per market.

Formally, the share-form dependent variable $y_{jm}\equiv \ln(s_{jm}/s_{0m})$ is constructed from observed market shares. In the nested logit model, the mapping from structural errors to these log-share ratios has a Jacobian that depends on nest size and the nesting parameter. Let $h_m(\mathbf e_m)$ denote the joint density of the structural errors in market $m$, and let $J_{gm}$ denote the number of products in nest $g$ in market $m$. The exact density of the observed log-share ratios is
\begin{equation}
\label{main_result_intro}
f_m(\mathbf y_m)
=
h_m\!\left(\mathbf e_m(\mathbf y_m)\right)
\prod_{g=1}^{G} (1-\sigma_g)^{J_{gm}-1}.
\end{equation}
The product term $\prod_{g=1}^{G}(1-\sigma_g)^{J_{gm}-1}$ is absent from standard share-equation estimators. Because it depends on $\sigma_g$, it enters the score for the nesting parameter and changes estimates of within-group substitution even when every market has the same menu. When choice sets vary, the term differs across markets, so the omission also misweights markets with different menus and ties the likelihood to product counts. These estimates feed directly into elasticities, markups, and welfare calculations.

The omission arises because standard share-equation estimation of nested logit uses the inverted share equation as an instrumental variables or GMM estimating equation rather than as a change of variables from structural errors to the observed log-share vector.   Once the likelihood is written for the observed log shares themselves, the Jacobian of that transformation must be included.  The count-instrument problem is separate. A count instrument is valid only if product counts are orthogonal to unobserved mean utility, and this fails when product counts enter demand directly, as they do when products crowd unobserved product space. The corrected likelihood connects these results: it identifies substitution without instrumenting the within-group share, allowing a direct product-count term to enter mean utility without using product counts as excluded instruments.

This variable-choice-set problem was identified, in different language, by \citet{ackerberg05}. They argued that standard logit-family models impose restrictions on how the unobserved product space expands with the number of products: each new product effectively occupies its own location in the latent space, so that additional products generate symmetric unobserved differentiation rather than crowding the existing alternatives. They proposed adding a parameterized function of product counts, $f(J;\gamma)$, to mean utility, with the free parameter $\gamma$ given a structural interpretation. Our contribution provides an econometric counterpart to their economic correction. Where \citet{ackerberg05} add product counts to utility as a modeling choice, we show that the exact nested-logit share-form likelihood requires a nest-size-dependent Jacobian term in the objective. This term is not a new preference parameter. It is a necessary consequence of writing the likelihood for the observed share equation.

The two corrections are complementary. The \citet{ackerberg05} correction adds a free parameter; ours adds none. Theirs is a behavioral assumption about the unobserved product space; ours is a statistical property of the share transformation that holds for any error distribution. Theirs shifts mean utility; ours modifies the score for $\sigma_g$. Most importantly, in settings where products genuinely crowd one another, the two corrections work best together. Our likelihood correction recovers the substitution parameter, and a parametric crowding term recovers the welfare effect of added variety.

The complementarity is not incidental. \citet{ackerberg05} document the bias
in nested-logit simulations, where the naive estimator misstates elasticities
by an order of magnitude and welfare gains from variety by large margins. But
their empirical implementation of the crowding correction is confined to the
simple logit, which has no within-nest share and therefore no within-share
instrument. In a nested logit setting, the count-based within-share instrument and the Ackerberg--Rysman crowding term rely on the same product-count variation. Once the crowding term is included in mean utility, product counts cannot also serve as excluded instruments for the within-share, so the Ackerberg--Rysman correction cannot be combined with standard count IV. We demonstrate this breakdown and our solution in the Medicare Advantage application. The corrected likelihood removes the within-share instrument from the problem, and with it the conflict. In short, the likelihood correction is what makes the Ackerberg--Rysman correction estimable in nested logit.

Product-count instruments are common in nested-logit and BLP-style demand estimation. In nested-logit applications, counts of rival products are often used to instrument the within-group share and appear prominently in best-practice guides and software packages \citep{conlon2020}. They are also widely used in practice. \citet{goldbergverboven2001} use counts of products in the same nest in estimating automobile demand. \citet{bjornerstedtverboven2016} use counts by subgroup, group, firm, and market in a merger analysis of Swedish analgesics, and \citet{millerweinberg2017} use product counts in evaluating the MillerCoors joint venture. Outside antitrust, \citet{donna2022} use the number of products within each distribution channel as an instrument for the within-group share in a nested-logit model of intermediary welfare.

Other work uses product-set instruments closely related to product counts. In
healthcare, \citet{town2003} instrument shares with out-of-market plan counts.
In radio, \citet{bew2016} instrument within-format shares with counts of
out-metro stations, on the grounds that stations earn most of their revenue at home,
so their presence in other markets is independent of taste shocks there.  These arguments address
correlation with unobserved quality, and they are sound in models where
product counts do not enter demand.   But when out-of-market products can
be consumed locally, as with radio signals, the
instrument counts products in the local choice set, so under crowding that
count enters demand directly no matter what determines entry. When
out-of-market products cannot be consumed locally, as with health plans
offered in other counties, exclusion fails if  out-of-market counts are
correlated with in-market counts.   This is the case in our Medicare Advantage
application, where out-of-market plan counts are strongly correlated with
in-market counts.

The exact likelihood has two further implications for practice. First, it eliminates the need to instrument for the within-nest share. The dependence between $\ln s_{j|g}$ and the structural error, which is the dependence that motivates within-share instruments in standard applications, is exactly what the Jacobian accounts for, so under the corrected likelihood the nesting parameter is identified without it. Price endogeneity is a separate matter and still requires instruments. Second, the correction is easy to adopt. The only change to a standard estimation routine is a single term in the objective function, $\sum_m \sum_g (J_{gm}-1)\ln(1-\sigma_g)$. Monte Carlo simulations confirm that absent crowding, the corrected likelihood recovers the substitution parameter without within-share instruments, and that when product crowding is present, the corrected estimator approaches the truth in cases where standard instrumental-variables estimation does not. We introduced the corrected likelihood in \citet{bakergeorge2024}, estimating
demand for local television news.  This remains, to our knowledge, the only implementation. The present paper develops the derivation, its consequences
for count instruments, and the evidence.  A Stata command implementing the corrected likelihood, \textit{exactnl}, accompanies
the paper.\footnote{Available at \url{https://github.com/lisa-george/exactnl} and installable in Stata via \texttt{net install exactnl, from("https://raw.githubusercontent.com/lisa-george/exactnl/main/")}.}

Medicare Advantage is a setting where policy acts directly on product variety. For contract year 2019, the Centers for Medicare and Medicaid Services eliminated the ``meaningful difference'' requirement that had previously limited the number of plans an insurer could offer in a market.\footnote{CMS introduced meaningful-difference review in response to duplicative plan offerings and beneficiary confusion. The 2010 Call Letter directed MA organizations to eliminate ``duplicative plan offerings that are not easily distinguished by beneficiaries and could cause beneficiary confusion.'' CMS further indicated that multiple offerings can ``confuse beneficiaries'' and announced that they expected organizations to offer no more than three plans per type in a market \citep{cms2010callletter}. Meaningful-difference review of bid submissions was formalized in subsequent annual guidance.} After the change, the number of plans available to a typical beneficiary roughly doubled over the following six years. The number of plans per county varies widely in the cross section and shifts sharply after the policy change, providing a setting in which the omitted Jacobian varies substantially across markets and count instruments fail if product counts enter mean utility. The central policy question, whether plan proliferation improves welfare through better matching or reduces welfare through duplication, depends on how the demand model attributes observed shares to genuine differentiation rather than to the mechanical effect of larger menus.

We present three main empirical findings. First, Monte Carlo simulations show that the corrected likelihood and a standard count-instrument estimator agree when the nested logit model is correctly specified and product counts do not enter demand, in other words absent crowding. Both estimators recover the true nesting parameter, and the corrected likelihood does so without an instrument for the within-nest share. In the no-crowding benchmark, the corrected likelihood recovers the true nesting parameter with smaller sampling variance.

Second, when product counts also shift mean utility, the standard count-instrument estimator attributes product-space crowding to within-group substitution.\footnote{We refer to this as the crowding case, recognizing that product counts can shift mean utility for reasons related to search, transaction costs or other factors.} The resulting bias grows with the importance of the product-count term, while the corrected likelihood remains close to the true substitution parameter. The simulations show why the difference between our corrected likelihood and count-instrument estimator is informative about the source of identifying variation.

Third, in our Medicare Advantage application, the count-instrument estimator
fails in a baseline specification with all plans in a single nest, producing a substitution parameter greater than one.   In a two-nest specification by plan type, the count
instruments produce substitution parameters more than twice the corrected
likelihood estimates. The estimated product-count term is negative, so count instruments are not valid in this setting. Corrected likelihood estimates imply that plan proliferation raised consumer surplus in most markets.
In 2024, ignoring crowding overstates the gains by 43 percent and implies no losing markets, while the preferred corrected estimates imply losses in two percent of markets, concentrated in the densest markets.  The direction of the crowding
effect is clear, but its magnitude is imprecise. 

The paper proceeds as follows. Section~\ref{sec:model} presents the model and derives the exact likelihood.  Section~\ref{sec:montecarlo} reports Monte Carlo simulations replicating and extending the experiments in \citet{ackerberg05}. Section~\ref{sec:application} presents the Medicare Advantage application. Section~\ref{sec:conclusion} concludes.

\section{Model} \label{sec:model}

\subsection{Setup}

We work with a two-level nested logit.  A researcher observes market shares of differentiated products
across markets $m = 1, \ldots, M$, which may vary across space, time, or both.
In each market a consumer $i$ chooses a product $j$ from group
$g \in \{1, \ldots, G\}$, or selects an outside option indexed $0$. The number
of products in group $g$, denoted $J_{gm}$, may differ across markets. 

Following \citet{berry94} and \citet{cardell1997}, consumer $i$'s utility from
product $j$ in group $g$ in market $m$ is
\begin{equation} \label{eq:utility}
u_{ijgm} = \delta_{jgm} + \nu_{igm}(\sigma_g) + (1 - \sigma_g)\,\varepsilon_{ijm},
\end{equation}
where $\delta_{jgm}$ is the mean utility of product $j$, the term
$\nu_{igm}(\sigma_g)$ is a group-specific component common to all products in
group $g$, and $\varepsilon_{ijm}$ is an idiosyncratic taste shock distributed
type-I extreme value. The parameter $\sigma_g \in [0,1)$ governs within-group
correlation: as $\sigma_g \to 1$, products within group $g$ become perfect
substitutes; when $\sigma_g = 0$, the model reduces to the standard logit. We
split mean utility into a systematic part and an unobserved part,
\begin{equation} \label{eq:meanutility}
\delta_{jgm} = \mu_{jgm} + \xi_{jm}, \qquad
\mu_{jgm} \equiv X_{jgm}\beta + \alpha\, p_{jm},
\end{equation}
where $X_{jgm}$ are observed characteristics, $p_{jm}$ is price, $\alpha<0$ is
the price coefficient, and $\xi_{jm}$ is unobserved product quality. Carrying $\mu_{jgm}$ and $\xi_{jm}$ separately
keeps the structural error visible in the estimating equation below, which is
where it plays its role.

Under these assumptions the market share of product $j$ takes the nested form
\begin{equation} \label{eq:shares}
s_{jm} = \underbrace{\frac{e^{\delta_{jgm}/(1-\sigma_g)}}
{\sum_{k \in J_{gm}} e^{\delta_{kgm}/(1-\sigma_g)}}}_{s_{j|g}} \cdot
\underbrace{\frac{D_{gm}^{1-\sigma_g}}{1 + \sum_{l} D_{lm}^{1-\sigma_l}}}_{s_{gm}},
\qquad
D_{gm} \equiv \sum_{k \in J_{gm}} e^{\delta_{kgm}/(1-\sigma_g)},
\end{equation}
The outside option $j=0$ is normalized in the usual way: $\delta_{0m}=0$, it
carries no structural error, and it sits in its own singleton nest, so it enters
only through the unit term in the denominator of (\ref{eq:shares}) and through
$s_{0m}$. With $s_{jm} = s_{j|g}\cdot s_{gm}$, \citet{berry94} showed that these
shares admit a linear estimating equation. Defining the log-share ratio
$y_{jm} \equiv \ln s_{jm} - \ln s_{0m}$, the Berry inversion yields
\begin{equation} \label{eq:berry}
y_{jm} = \mu_{jgm} + \sigma_g \ln s_{j|g} + \xi_{jm}.
\end{equation}
The dependent variable $y_{jm}$ is observed. The right-hand side contains the
systematic mean utility $\mu_{jgm}$, the within-group share $\ln s_{j|g}$, and
the structural error $\xi_{jm}$. Standard practice estimates (\ref{eq:berry}) as
a linear equation in which $\xi_{jm}$ is the only stochastic object,
instrumenting for price and, separately, for the within-group share. Product counts are common instruments for the within-group share, so we refer to
this throughout as the \emph{count IV estimator}, to distinguish it both
from our corrected likelihood estimator and from the uncorrected Gaussian likelihood.

This estimating equation is used widely in the industrial organization literature, and it is valid as a linear moment condition. A likelihood for the observed log shares must do more. The dependent variable $y_{jm}$ is a transformation of the observed shares, and the shares are the model's endogenous outcome. A transformed random variable has a density that
differs from the density of the underlying error by a Jacobian factor, which we derive in the next subsection. 

\subsection{The Exact Likelihood}

The dependent variable $y_{jm}$ is a transformation of the structural errors
$\xi_{jm}$. Let $\mathcal J_{gm}$ denote the set of inside products in nest $g$
in market $m$, let $J_{gm}=|\mathcal J_{gm}|$, and let $\mathcal G_m$ denote the
set of nonempty inside nests in market $m$. For a product $j\in\mathcal J_{gm}$,
write $y_{jgm}\equiv y_{jm}$. If the density of the structural-error vector
$\mathbf e_m=(\xi_{jgm}:g\in\mathcal G_m,\ j\in\mathcal J_{gm})'$ is
$h_m(\mathbf e_m)$, then the density of the transformed vector
$\mathbf y_m=(y_{jgm}:g\in\mathcal G_m,\ j\in\mathcal J_{gm})'$ is given by the
change-of-variables formula
\begin{equation} \label{eq:changeofvar}
f_m(\mathbf y_m)
=
h_m\!\left(\mathbf e_m(\mathbf y_m)\right)
\left| \det \frac{\partial \mathbf e_m}{\partial \mathbf y_m} \right|
=
h_m\!\left(\mathbf e_m(\mathbf y_m)\right)
\left| \det \frac{\partial \mathbf y_m}{\partial \mathbf e_m} \right|^{-1}.
\end{equation}
The task is to compute the Jacobian and its determinant. For $\sigma_g<1$ and
strictly positive shares this transformation is one-to-one: the determinant
computed below is nonzero, so the map $\mathbf e_m \mapsto \mathbf y_m$ is
locally invertible, and the Berry inversion supplies the global inverse used in
estimation.

Rewriting (\ref{eq:berry}) by substituting the expression for the within-group
share in terms of the structural primitives, and writing
$\delta_{jgm}=\mu_{jgm}+\xi_{jgm}$, we can express $y_{jgm}$ as
\begin{equation} \label{eq:yexpanded}
y_{jgm}
=
\frac{1}{1-\sigma_g}\left(\mu_{jgm}+\xi_{jgm}\right)
-
\sigma_g
\ln\!\left(
\sum_{k\in\mathcal J_{gm}}
\exp\left\{\frac{\mu_{kgm}+\xi_{kgm}}{1-\sigma_g}\right\}
\right).
\end{equation}
Differentiating (\ref{eq:yexpanded}) with respect to $\xi_{kgm}$, and using the
within-group share
\[
s_{k\mid g,m}
=
\frac{
\exp\left\{(\mu_{kgm}+\xi_{kgm})/(1-\sigma_g)\right\}
}{
\sum_{\ell\in\mathcal J_{gm}}
\exp\left\{(\mu_{\ell gm}+\xi_{\ell gm})/(1-\sigma_g)\right\}
},
\]
gives the partial derivatives
\begin{equation} \label{eq:partials}
\frac{\partial y_{jgm}}{\partial \xi_{kgm}}
=
\begin{cases}
\dfrac{1}{1-\sigma_g}
-
\dfrac{\sigma_g}{1-\sigma_g}\,s_{k\mid g,m}
& \text{if } j=k, \\[8pt]
-\dfrac{\sigma_g}{1-\sigma_g}\,s_{k\mid g,m}
& \text{if } j\neq k,\ j,k\in\mathcal J_{gm}, \\[8pt]
0
& \text{if } j \text{ and } k \text{ are in different nests.}
\end{cases}
\end{equation}
The cross-nest zero derivatives imply that the full Jacobian is block diagonal,
so its determinant factors across nests. Let
$\mathbf A_{gm}$ denote the $J_{gm}\times J_{gm}$ block corresponding to nest
$g$ in market $m$. This block has the form
\begin{equation} \label{eq:blockJ}
\mathbf A_{gm}
=
\frac{1}{1-\sigma_g}
\left(
\mathbf I_{J_{gm}}
-
\sigma_g\,\mathbf 1_{J_{gm}}\mathbf s_{gm}'
\right),
\end{equation}
where $\mathbf s_{gm}=(s_{k\mid g,m}:k\in\mathcal J_{gm})'$ and
$\mathbf 1_{J_{gm}}$ is a $J_{gm}$-vector of ones. By the matrix determinant
lemma,
\[
\left|\det \mathbf A_{gm}\right|
=
\left(\frac{1}{1-\sigma_g}\right)^{J_{gm}}
\left(1-\sigma_g\,\mathbf s_{gm}'\mathbf 1_{J_{gm}}\right).
\]
Because the within-nest shares sum to one,
$\mathbf s_{gm}'\mathbf 1_{J_{gm}}=1$, so
\[
\left|\det \mathbf A_{gm}\right|
=
\left(\frac{1}{1-\sigma_g}\right)^{J_{gm}}
(1-\sigma_g)
=
(1-\sigma_g)^{-(J_{gm}-1)}.
\]
It follows that
\[
\left|
\det \frac{\partial \mathbf y_m}{\partial \mathbf e_m}
\right|
=
\prod_{g\in\mathcal G_m}
(1-\sigma_g)^{-(J_{gm}-1)}.
\]
Substituting into (\ref{eq:changeofvar}), the exact density of the observed
share-form dependent variable is
\begin{equation} \label{eq:mainresult}
\boxed{
f_m(\mathbf y_m)
=
h_m\!\left(\mathbf e_m(\mathbf y_m)\right)
\prod_{g\in\mathcal G_m}
(1-\sigma_g)^{J_{gm}-1}
}.
\end{equation}

Equation (\ref{eq:mainresult}) is our main result. The exact likelihood of the
Berry-inversion dependent variable is the density of the implied structural
errors multiplied by a Jacobian factor that depends on the nesting parameters
$\sigma_g$ and the nest sizes $J_{gm}$. This factor is not part of standard
share-equation estimators. Because it depends on $\sigma_g$, it enters the
score for the nesting parameter. When $J_{gm}$ varies across markets, the
omitted term also changes how markets with different choice sets contribute to
the likelihood, and omitting it changes the value of $\sigma_g$ that maximizes
the likelihood. These estimates feed directly into substitution patterns,
elasticities, markups, and welfare calculations.

\subsection{Properties} \label{sec:properties}

Equation~(\ref{eq:mainresult}) has several features that structure the rest of
the paper. The first two are important enough to separate and state formally:  the likelihood correction is implied by the change of
variables and does not depend on the distributional form of the structural errors,
whereas any Gaussian implementation is a separate, stronger assumption.

\begin{proposition}[Distribution-free likelihood correction]
\label{prop:distfree}
The factor
\[
\prod_{g\in\mathcal G_m}(1-\sigma_g)^{J_{gm}-1}
\]
in (\ref{eq:mainresult}) is implied by the change of variables from structural
errors to observed log-share ratios. It holds for any absolutely continuous
density $h_m(\cdot)$ of the structural errors. It introduces no additional free
parameters: the factor is fixed by the nest sizes $J_{gm}$ and the nesting
parameters $\sigma_g$.
\end{proposition}

\begin{proposition}[Gaussian implementation]
\label{prop:gaussian}
If the structural errors are correctly specified as homoskedastic normal, then
the corrected likelihood in (\ref{eq:mainresult}) is the exact likelihood, and
its maximizer is the maximum likelihood estimator of the demand parameters,
together with any variance parameters. If the error distribution is misspecified,
the maximizer is a quasi-maximum likelihood estimator: it converges to the
pseudo-true parameter that maximizes the expected corrected Gaussian
log-likelihood. Inference then requires a sandwich or bootstrap covariance
estimator. Whether the pseudo-true nesting parameter equals the structural
nesting parameter depends on the corresponding quasi-score conditions, not on
correct specification of the conditional mean alone.
\end{proposition}

The separation of these two propositions is important in the remainder of the
paper. The likelihood correction is not a modeling assumption; it is an
accounting identity for the transformation from structural errors to observed
shares (Proposition~\ref{prop:distfree}). What a researcher assumes about the
error distribution is a distinct choice, and the consequences are the ordinary
ones for quasi-likelihood \citep{white1982, gourieroux1984} (Proposition~\ref{prop:gaussian}). In particular, the
corrected Gaussian likelihood is exact under normality and a quasi-likelihood
otherwise. Under misspecification, curvature-based standard errors are
unreliable, so inference should use clustered sandwich or bootstrap standard
errors. We examine the robustness of $\hat\sigma_g$ to non-normal errors in the
simulations rather than claiming it generally.

We describe the remaining properties of equation~(\ref{eq:mainresult}) more
briefly.

\emph{Parameter bounds.} The likelihood correction enters the log-likelihood as
$(J_{gm}-1)\ln(1-\sigma_g)$ for nest $g$ in market $m$. This term diverges to
$-\infty$ as $\sigma_g\to 1$, penalizing values of $\sigma_g$ near the upper
boundary, increasingly so in markets with larger nests. The penalty does not
replace the parameter-space restriction $\sigma_g\in[0,1)$. In particular, it
does nothing at the lower bound, but it makes the upper boundary costly to
approach.

\emph{Nesting-structure.} Because the likelihood correction
$(J_{gm}-1)\ln(1-\sigma_g)$ becomes more negative as either  $J_{gm}$ or $\sigma_g$ increase, the corrected log-likelihood penalizes high values of $\sigma_g$ more strongly in larger nests.    A nest of 25 products with $\sigma_g=0.7$ contributes 
$24\times\ln(0.3)\approx -28.9$ to the log-likelihood in that market; a nest of
5 products at the same $\sigma_g$ contributes only $-4.8$. The uncorrected
share-form likelihood imposes no such penalty. Comparisons of alternative nesting structures therefore include the different Jacobian contributions implied by their nest sizes. We use such comparisons as a diagnostic rather than as a formal model-selection result, but find it to be a practical consequence of the correction's dependence
on nest size.

\emph{Relationship to mean utility correction.} Our likelihood
correction addresses a distinct channel that arises in the same empirical
environments considered by \citet{ackerberg05}, namely settings in which the number of
products varies across markets. Their concern is broader than ours. They argue
that standard logit-family models, including logit, nested logit, and
random-coefficients specifications, impose restrictions on how the unobserved
product space expands with the number of products, restrictions that can bias
elasticities and welfare. Our result is narrower but exact: it corrects the
share-form likelihood for the nested logit.

The two corrections are distinct in several respects. First,
\citeauthor{ackerberg05} introduce a parameterized function of product counts,
with examples including $\ln(\gamma/J+1-\gamma)$ and a log-linear form
$\gamma\ln J$, where $\gamma$ is a free parameter. Our likelihood correction
introduces no additional free parameter, since the correction is fixed by
$J_{gm}$ and $\sigma_g$. Second, their parametric correction is a behavioral model of how
unobserved product space scales with the number of products; ours is a
statistical consequence of the share transformation that holds for any
absolutely continuous error distribution. Third, the two enter the estimator at
different points: the parametric correction shifts mean utility, while
the likelihood correction adds a term to the log-likelihood and modifies the
score for $\sigma_g$. Fourth, the two are complementary rather than competing.
Our likelihood correction addresses estimation of within-nest substitution in
the share-form likelihood. Their parametric correction is
needed to measure the direct utility and welfare consequences of product
congestion, considered in the next subsection.

\subsection{Product Congestion, Utility and Welfare}
\label{sec:saturation}

Following \citet{ackerberg05}, we allow the number of products in a market to enter mean utility directly and consider how the parametric correction  coexists with the likelihood
correction.

Let mean utility include a term in the nest's product count:
\begin{equation} \label{eq:util_sat}
u_{ijgm}
=
x_{jgm}\beta
+
\alpha p_{jgm}
+
\tau_g r(J_{gm})
+
\xi_{jgm}
+
\nu_{igm}(\sigma_g)
+
(1-\sigma_g)\varepsilon_{ijm},
\end{equation}
where $r(\cdot)$ is a known increasing function, for example $r(J)=\ln J$, and
$\tau_g$ indexes product-space congestion in nest $g$. The case $\tau_g=0$
corresponds to the standard model with no additional mean-utility effect of nest
size. The case $\tau_g<0$ means that larger nests lower the utility of products
in the nest, consistent with congestion, search costs or other demand-side costs of larger choice sets. 

Because $\tau_g r(J_{gm})$ is common to every product in nest $g$ and market $m$,
it passes through the Berry inversion like any other component of mean utility.
The estimating equation becomes
\begin{equation} \label{eq:berry_sat}
\ln s_{jgm} - \ln s_{0m}
=
x_{jgm}\beta
+
\alpha p_{jgm}
+
\sigma_g \ln s_{j\mid g,m}
+
\tau_g r(J_{gm})
+
\xi_{jgm}.
\end{equation}

Equations~(\ref{eq:mainresult}) and (\ref{eq:berry_sat}) show that the product count $J_{gm}$ plays two
distinct roles:
\begin{itemize}
\item \emph{Likelihood channel.} The term
$(J_{gm}-1)\ln(1-\sigma_g)$ enters the log-likelihood through the change of
variables. It is statistical, introduces no additional free parameter, holds for
any absolutely continuous error distribution, and modifies the score for
$\sigma_g$. It is present whether or not products crowd one another.

\item \emph{Utility channel.} The term $\tau_g r(J_{gm})$ enters
mean utility. It is behavioral, carries the free parameter $\tau_g$, and is
present only if added products change utility directly through product-space
congestion or related demand-side mechanisms.
\end{itemize}

The two channels are conceptually distinct but empirically connected because both depend on  $J_{gm}$.  The entanglement is strongest when products within a nest are similar:  holding the nest's total share fixed, adding products mechanically reduces each product's conditional share, so  $s_{j\mid g,m}$ is approximately
$1/J_{gm}$ and $\ln s_{j\mid g,m}$ moves approximately with $-\ln J_{gm}$.  The
likelihood correction estimates $\sigma_g$ from the share-form likelihood without
requiring a count instrument for the within-group share, but it does not identify
the mean-utility congestion parameter $\tau_g$. Recovering the welfare
consequences of product-space congestion requires estimating $\tau_g$ in mean
utility. This is why we describe the likelihood correction and the
parametric utility correction as complementary.

We discuss the difficulty of separating $\sigma_g$ from $\tau_g$ when
$\ln s_{j\mid g,m}$ moves closely with $\ln J_{gm}$ in Section~\ref{sec:identification}. In the Monte Carlo simulations in
Section~\ref{sec:montecarlo}, we use $r(J)=\ln J$, so that the congestion design
$-\sigma_g\gamma\ln J_{gm}$ is exactly $\tau_g r(J_{gm})$ with
$\tau_g=-\sigma_g\gamma$.

\subsection{The Within-Group Share and the Role of Instruments}
\label{sec:withinshare}

A central practical consequence of our likelihood correction concerns the within-group
share $\ln s_{j|g}$. In the standard estimating equation (\ref{eq:berry}),
$\ln s_{j|g}$ appears as a regressor and is mechanically correlated with
$\xi_{jm}$.  Product $j$'s within-nest share depends on the structural errors of
all products in the nest, including its own. Typical practice treats this as an
endogeneity problem and instruments for $\ln s_{j|g}$, often with the count or 
characteristics of rival products in the nest.

The exact likelihood gives a different account of the same dependence. The
correlation between $\ln s_{j|g}$ and $\xi_{jm}$ is not a feature of the world
that requires an instrument; it is induced by the share transformation, and it is
exactly what the Jacobian factor accounts for. Under the corrected likelihood,
$\sigma_g$ is identified from the joint density of observed shares without an
instrument for the within-group share. Instruments remain necessary for
endogeneity that arises from correlation with unobserved product quality, most importantly for price.

This has two implications for instrument choice.  First, instruments for $\ln s_{j|g}$ are no longer needed: the
mechanical dependence between the within-share and $\xi_{jm}$ created by the share
transformation is handled by the Jacobian, so count instruments are unnecessary. Second, in environments with crowding or other features where product counts enter
mean utility directly, product counts fail the usual orthogonality
condition, so count instruments are invalid.  Unnecessary in the first case and invalid in the second, they are no longer the right excluded instruments.  This is
the same point \citet{ackerberg05} made conceptually: identification in the
standard approach can come from variation in product counts rather than from prices
or characteristics.  Valid instruments for the corrected model are cost shifters
for price and possibly
differentiation-based instruments in the spirit of \citet{gandhi2020}.

\subsection{Identification}
\label{sec:identification}

We write the demand equation as
\begin{equation}
y_{jgm}
=
x_{jgm}\beta
+
\alpha p_{jgm}
+
\sigma_g \ln s_{j\mid g,m}
+
\tau_g r(J_{gm})
+
\xi_{jgm},
\label{eq:identification_equation}
\end{equation}
where $r(J_{gm})$ is a known function of the number of products in nest $g$ and market $m$, for example $r(J)=\ln J$. The likelihood correction adds
\begin{equation}
\sum_m\sum_g (J_{gm}-1)\ln(1-\sigma_g)
\label{eq:identification_jacobian}
\end{equation}
to the objective function. Identification therefore involves three distinct objects: the standard demand parameters $(\beta,\alpha)$, the nesting parameter $\sigma_g$, and the product-count utility parameter $\tau_g$.

\paragraph{Product characteristics and price.}
The coefficients on observed product characteristics are identified from variation in $x_{jgm}$ that is excluded from unobserved product quality $\xi_{jgm}$. Price remains a standard endogeneity problem: if $p_{jgm}$ is correlated with $\xi_{jgm}$, then $\alpha$ requires excluded cost or payment-side instruments. Section~\ref{sec:scores} shows how such instruments enter the corrected framework through a control function.

\paragraph{The nesting parameter.}
In the standard linear IV approach, $\sigma_g$ is identified from instrumented variation in $\ln s_{j\mid g,m}$. Product-count instruments are attractive because the number of products in a nest mechanically affects within-nest shares. This source of variation is not valid, however, when product counts also enter mean utility. If the true equation contains $\tau_g r(J_{gm})$ but the estimated equation omits it, the structural residual is
\begin{equation}
u_{jgm}
=
\tau_g r(J_{gm})
+
\xi_{jgm}.
\end{equation}
A count instrument $Z_{gm}=Z(J_{gm})$ is valid only if
\begin{equation}
E\!\left[
Z_{gm}
\left(
\tau_g r(J_{gm})+\xi_{jgm}
\right)
\right]
=
0.
\label{eq:count_iv_failure}
\end{equation}
This moment generally fails because the instrument and the omitted utility term are functions of the same product count. The resulting bias is not a failure of the instrumental variable approach itself, but the consequence of using product counts as excluded instruments while omitting a product-count utility term.

Including $\tau_g r(J_{gm})$ in the linear IV equation as recommended by \cite{ackerberg05} removes the omitted-variable channel, but it also removes the clean exclusion argument for count instruments because they appear directly in mean utility. The same count variation cannot credibly identify $\tau_g$ as an included utility shifter and also serve as excluded variation for $\sigma_g$. If the excluded instrument is the same function of $J_{gm}$ as the included term, the rank condition fails. If the excluded instrument is a different function of $J_{gm}$, identification of $\sigma_g$ rests on functional-form differences among product-count functions rather than on an economic exclusion restriction. 

The corrected likelihood identifies $\sigma_g$ differently. It treats $\ln s_{j\mid g,m}$ as part of the transformation from structural errors to observed shares and uses the joint density of the share vector. The estimate of $\sigma_g$ is disciplined by the residual fit of the share equation and by the Jacobian term in \eqref{eq:identification_jacobian}. Product counts therefore need not serve as excluded instruments for the within-nest share. This allows product-count variation to enter mean utility directly  without also being used as an excluded instrument for substitution.

\paragraph{The utility parameter.}
The parameter $\tau_g$ is identified from variation in $r(J_{gm})$ that shifts mean utility, conditional on the other included covariates:
\begin{equation}
E\!\left[r(J_{gm})\xi_{jgm}\right]=0.
\end{equation}
It is therefore a mean-utility parameter, not a likelihood parameter. The likelihood correction alone does not identify $\tau_g$ unless the term $\tau_g r(J_{gm})$ is included in the demand equation. This is a substantive restriction, since entry is endogenous: firms add products where demand is strong, making $r(J_{gm})$ positively correlated with $\xi_{jgm}$. The correlation biases $\hat\tau_g$ upward, so a negative estimate is less negative than the truth. Estimates of $\tau_g<0$ are therefore conservative.

The preceding discussion of $\sigma_g$ shows why this creates a problem for count-IV specifications. If $\tau_g r(J_{gm})$ is omitted, count instruments for $\ln s_{j\mid g,m}$ are correlated with the omitted product-count utility term and $\sigma_g$ is biased. If $\tau_g r(J_{gm})$ is included, product counts are no longer excluded instruments; they are regressors in mean utility. A count-IV specification can then identify both $\sigma_g$ and $\tau_g$ only with additional non-count instruments for the within-nest share, or by relying on functional-form differences among alternative functions of $J_{gm}$. The corrected likelihood avoids this conflict because $\sigma_g$ is identified from the share-form likelihood rather than from count instruments, leaving product-count variation available to identify $\tau_g$ in mean utility.

The welfare implication follows directly. Because $\tau_g r(J_{gm})$ enters mean utility, welfare calculations that evaluate changes in product variety require an estimate of $\tau_g$, or an explicit maintained assumption that $\tau_g=0$. The likelihood correction does not by itself recover the welfare consequences of congestion, but in avoiding the need for share instruments allows identification of  $\tau_g$ directly through a parametric control.  

\paragraph{Panel data and fixed effects.}
Panel data can provide additional variation in product counts, but it does not change the underlying distinction between count-IV and likelihood identification. Let tildes denote variables residualized with respect to market and time period effects. The residualized demand equation is
\begin{equation}
\widetilde y_{jgm}
=
\widetilde x_{jgm}\beta
+
\alpha \widetilde p_{jgm}
+
\sigma_g \widetilde{\ln s}_{j\mid g,m}
+
\tau_g \widetilde{r(J_{gm})}
+
\widetilde \xi_{jgm}.
\label{eq:residualized_identification}
\end{equation}
If $\tau_g r(J_{gm})$ is omitted, count IV requires
\begin{equation}
E\!\left[
\widetilde Z_{gm}
\left(
\tau_g \widetilde{r(J_{gm})}
+
\widetilde \xi_{jgm}
\right)
\right]
=
0.
\label{eq:residualized_count_iv_failure}
\end{equation}
Thus panel fixed effects solve the count-IV problem only when they absorb the relevant product-count variation. If product counts vary within markets over time, or across nests within market-period cells, residual product-count variation remains. When that residual variation is used as a count instrument for the within-nest share, it remains correlated with the omitted residual congestion term unless $\tau_g=0$.

Panel variation can identify $\tau_g$ when $r(J_{gm})$ varies after residualizing by the included fixed effects. In that case, a corrected-likelihood specification can use residual product-count variation to estimate $\tau_g$, while $\sigma_g$ is identified by the likelihood rather than by count instruments. A count-IV specification faces the same conflict as in the cross-sectional case: if $\tau_g r(J_{gm})$ is omitted, $\sigma_g$ is biased; if it is included, product counts are regressors and cannot also provide a credible excluded instrument for $\ln s_{j\mid g,m}$ without additional non-count instruments or reliance on functional-form differences among count measures.

\subsection{Log-Likelihood, Score Functions, and Implementation} \label{sec:scores}

For estimation, suppose the structural errors are i.i.d.\ normal with mean zero
and variance $\omega^2$, and write the residual from the Berry inversion as
\[
r_{jgm} \equiv y_{jgm} - X_{jgm}\beta - \alpha\, p_{jm} - \sigma_g \ln s_{j|g}
- \tau_g\, r(J_{gm}),
\]
where $\tau_g=0$ recovers the specification without a product-count term.
Throughout we treat observed shares as exact choice probabilities, as is standard for log-share models estimated on administrative enrollment data.

The log-likelihood contribution of observation $j$ in market $m$ is
\begin{equation} \label{eq:loglik}
\ell_{jgm} = -\frac{r_{jgm}^2}{2\omega^2} - \frac{1}{2}\ln(2\pi\omega^2)
+ \frac{J_{gm} - 1}{J_{gm}}\ln(1-\sigma_g).
\end{equation}
The first two terms are the standard normal log-likelihood of the residual. The
third is the Jacobian correction, spread equally over the $J_{gm}$ products in
the group so that summing over the group recovers the full correction
$(J_{gm}-1)\ln(1-\sigma_g)$.\footnote{An equivalent implementation would apply the full
correction once per market-nest group rather than distributing it across
observations.}

Differentiating (\ref{eq:loglik}) yields the score functions
\begin{align}
\frac{\partial \ell_{jgm}}{\partial \beta}
&= \frac{X_{jgm}\, r_{jgm}}{\omega^2}, \label{eq:score_beta} \\[6pt]
\frac{\partial \ell_{jgm}}{\partial \alpha}
&= \frac{p_{jm}\, r_{jgm}}{\omega^2}, \label{eq:score_alpha} \\[6pt]
\frac{\partial \ell_{jgm}}{\partial \tau_g}
&= \frac{r(J_{gm})\, r_{jgm}}{\omega^2}, \label{eq:score_tau} \\[6pt]
\frac{\partial \ell_{jgm}}{\partial \omega}
&= \frac{r_{jgm}^2}{\omega^3} - \frac{1}{\omega}, \label{eq:score_omega} \\[6pt]
\frac{\partial \ell_{jgm}}{\partial \sigma_g}
&= \frac{(\ln s_{j|g})\, r_{jgm}}{\omega^2}
- \frac{J_{gm} - 1}{J_{gm}}\cdot\frac{1}{1-\sigma_g}. \label{eq:score_sigma}
\end{align}
The scores for the mean-utility parameters $\beta$, $\alpha$, and $\tau_g$ take
the standard form, as does the score for the error standard deviation $\omega$.
The score for the nesting parameter $\sigma_g$ is where the correction enters: the
second term, $-\frac{J_{gm}-1}{J_{gm}}\frac{1}{1-\sigma_g}$, is absent in the
standard estimator. It is negative and larger in magnitude in high-$J_{gm}$ nests,
so holding residual fit fixed it pushes the estimate toward lower $\sigma_g$ in
markets with many products. This offsets the residual-fit term, which pushes the
other way. In a large nest the conditional shares $s_{j\mid g,m}$ are small and
widely dispersed, and fitting that dispersion calls for a higher $\sigma_g$. The
correction  prevents nest size alone from driving
$\sigma_g$ toward its upper bound, and it does so most where nests are largest.
This is the score-level counterpart of the nesting-structure discipline described
in Section~\ref{sec:properties}. The realized estimate reflects the balance between
the two terms.

The structure of the likelihood makes estimation simple. Given the nesting
parameters, the demand equation is linear, so the mean-utility parameters and
the error variance can be concentrated out by least squares. Up to a constant,
the profile log-likelihood is
\begin{equation} \label{eq:profile}
\ell^{p}(\sigma_1,\ldots,\sigma_G)
= -\frac{N}{2}\,\ln\!\left(\frac{SSR(\sigma)}{N}\right)
+ \sum_{g}\Bigl[\,\sum_{m}(J_{gm}-1)\Bigr]\ln(1-\sigma_g),
\end{equation}
where $SSR(\sigma)$ is the sum of squared residuals from the linear fit at the
candidate nesting parameters. Estimation reduces to a low-dimensional search
over $\sigma_g\in[0,1)$, with every other parameter recovered by least squares
at the maximum.

When price is correlated with unobserved quality $\xi_{jm}$, the score for
$\alpha$ in (\ref{eq:score_alpha}) is no longer a valid moment condition, and
price requires excluded instruments as in any demand model. Our implementation
uses a control function. A first stage regresses price on the excluded
instruments, the exogenous regressors including the product-count term, and the
fixed effects; the first-stage residual then enters the structural equation as
an additional regressor, absorbing the endogenous component of $\xi_{jm}$. The
within-group share $\ln s_{j|g}$ requires no instrument at any stage.

The companion Stata command \textit{exactnl} implements this estimator: a
control-function first stage, high-dimensional fixed effects, and the profile
likelihood (\ref{eq:profile}) maximized by grid search over the nesting
parameters. The command reports the boundary likelihood-ratio test of
$\sigma_g=0$ against the conservative \citet{koddepalm1986} critical values and
computes cluster block-bootstrap standard errors for $\sigma_g$ and $\tau_g$,
which we use for inference in the application. An option removes the Jacobian
term, reproducing the uncorrected likelihood for comparison.

\bigskip

%\hrule
%\bigskip

\section{Monte Carlo Simulations} \label{sec:montecarlo}

The simulations below compare our corrected likelihood estimator to a linear estimator  that instruments the within-nest share with the count of rival products in the nest.
We label this Count IV.  We compare the estimators across two
environments, one in which the nested logit is correctly specified, and one with unmodeled product congestion.

\subsection{Design}

We simulate a market with one product nest and an outside good with a true nesting parameter of
$\sigma = 0.5$, a price coefficient $\alpha = -1$, and a single product characteristic with coefficient $\beta = 1$. The structural error is drawn i.i.d.\
normal. Each replication contains $M = 1{,}000$ markets, and we report averages
over $R = 200$ replications, so that estimation error is small and the reported
figures are close to probability limits. The number of products $J_m$ varies
across markets.  We draw it three ways: 
 uniform on $[1,10]$, uniform on $[1,30]$, and a
distribution calibrated to our Medicare Advantage data of
Section~\ref{sec:application} to confirm that the results do not depend on the shape of the choice-set distribution.

We introduce congestion following  \citet{ackerberg05}.  With congestion, the number of
products enters mean utility through the term $-\sigma\gamma\ln J_m$, where
$\gamma$ indexes the strength of congestion. At $\gamma = 0$ there is no
congestion and the nested logit is correctly specified.  At $\gamma = 1$ each
additional product fully crowds the existing ones. In every cell we run both
estimators on the same simulated data.  Importantly, in the simulations,  neither estimator is
given a congestion correction, so that the congestion column  measures the bias
each estimator incurs when the environment contains congestion that the estimator ignores.

\subsection{Simulations without Congestion}

Table~\ref{tab:sim_nocong} reports the estimated nesting parameter when the nested
logit is correctly specified ($\gamma = 0$). Across all three variable choice sets, count IV and the corrected likelihood recover the true
$\sigma = 0.5$ and agree with each other to three or four decimal places. The mean
gap between the two estimators never exceeds $0.0004$.  

\begin{table}[htbp]
\centering
\caption{Simulations Without Congestion}
\label{tab:sim_nocong}
\begin{tabular}{lccccc}
\toprule
 & \multicolumn{2}{c}{$\hat\sigma$} & Mean gap & \multicolumn{2}{c}{RMSE} \\
\cmidrule(lr){2-3}\cmidrule(lr){5-6}
$J$ distribution & Count IV & \shortstack{Corrected\\Likelihood} & (IV $-$ corr.) & Count IV & \shortstack{Corrected\\Likelihood}\\
\midrule
$U[1,10]$     & 0.5006 & 0.5002 & $\phantom{-}$0.0003 & 0.0093 & 0.0047 \\
$U[1,30]$     & 0.4994 & 0.4997 & $-$0.0003           & 0.0067 & 0.0036 \\
MA-calibrated & 0.5002 & 0.4998 & $\phantom{-}$0.0004 & 0.0063 & 0.0034 \\
\bottomrule
\end{tabular}
\par\smallskip
\footnotesize\begin{minipage}{\textwidth}
\emph{Notes.} True $\sigma = 0.5$; $M = 1000$ markets; $R = 200$
replications. Each row shows one nest-size ($J$) distribution. The Count IV
estimator instruments the within-nest share with the rival count; the
corrected likelihood uses no within-share instrument. ``Mean gap'' is
the mean of $\hat\sigma$(Count IV) $-$ $\hat\sigma$(Corrected Likelihood) across
replications, computed from the unrounded estimates; it therefore need not equal
the difference of the two rounded columns shown. Absent congestion the two
estimators agree to within Monte Carlo noise across all three $J$ distributions.
\end{minipage}
\end{table}

This agreement is the first result.  When the
nested logit is the true model, our likelihood  correction   recovers the same $\sigma$ that count  IV
recovers without an instrument for the within-nest share. The agreement does not
depend on the choice-set distribution, it holds whether nest sizes range over ten
products or thirty, and whether they follow a uniform draw or the empirical
Medicare Advantage shape. Our correction is in this sense a safe default.  When the nested logit is correctly specified and product counts do not affect mean utility, the two estimators coincide.  The corrected likelihood is also more efficient, with a root mean squared error about half that of count IV.  This is expected. Under correct specification, the corrected likelihood is the maximum likelihood estimator, and the maximum-likelihood estimator is asymptotically efficient under standard regularity conditions. The estimator extracts information in the within-nest shares directly rather than through an instrument. The efficiency advantage will matter most in applications with few markets or few products per nest.

\subsection{Simulations with Congestion}

The two estimators diverge when congestion is added to the simulation.   
Table~\ref{tab:sim_cong} holds the choice-set distribution at the
shape calibrated to our Medicare Advantage application (Section~\ref{sec:application}) and raises the congestion parameter $\gamma$ from zero
to one.  The true nesting parameter remains 
$\sigma = 0.5$.    Bias in the count IV grows steadily with congestion.  At $\gamma=1$, $\sigma$ is biased upward by $0.24$, nearly half  the true value.  Importantly, the bias in the corrected likelihood also grows with congestion, but modestly.  At full congestion the bias is $0.0598$.  

\begin{table}[htbp]
\centering
\caption{Simulations With Congestion}
\label{tab:sim_cong}
\begin{tabular}{cccccc}
\toprule
 & \multicolumn{2}{c}{Count IV} & \multicolumn{2}{c}{Corrected Likelihood} & Mean gap \\
\cmidrule(lr){2-3}\cmidrule(lr){4-5}
$\gamma$ & $\hat\sigma$ & Bias & $\hat\sigma$ & Bias & (IV $-$ corr.) \\
\midrule
0.00 & 0.5002 & $\phantom{-}$0.0002 & 0.4998 & $-$0.0002           & 0.0004 \\
0.27 & 0.5653 & $\phantom{-}$0.0653 & 0.5209 & $\phantom{-}$0.0209 & 0.0444 \\
0.50 & 0.6208 & $\phantom{-}$0.1208 & 0.5361 & $\phantom{-}$0.0361 & 0.0847 \\
1.00 & 0.7418 & $\phantom{-}$0.2418 & 0.5598 & $\phantom{-}$0.0598 & 0.1820 \\
\bottomrule
\end{tabular}
\par\smallskip
\footnotesize\begin{minipage}{\textwidth}
\emph{Notes.} True $\sigma = 0.5$; MA-calibrated $J$ distribution;
$M = 1000$ markets, $R = 200$ replications. Congestion enters mean
utility as $-\sigma\gamma\ln J$. Neither estimator includes a
congestion correction.
\end{minipage}
\end{table}

The difference in the estimators under congestion arises from the count instrument's exclusion restriction.  Count IV identifies $\sigma$ from the within-nest
share, instrumented by the rival count. Under congestion, the omitted term
$-\sigma\gamma\ln J_m$ is a function of the number of products, and so is the
instrument; the instrument is therefore correlated with the part of the error the
estimator has left out, and the exclusion restriction fails. The bias grows
with $\gamma$, reaching $0.24$ at full congestion, nearly half the true
parameter. The corrected likelihood does not use a product-count instrument.  Its misspecification enters only
through the fit of the residual, and the resulting bias stays below $0.06$
across the full range of $\gamma$. The pattern in
Table~\ref{tab:sim_cong} follows: as congestion strengthens, the IV estimate
pulls steadily away from the truth while the corrected estimate stays close.

We argue that this pattern supports adoption of the corrected likelihood.   A researcher cannot know in advance whether the products in their data crowd one another. If they do
not, the two estimators agree.  But if congestion is present, count IV is biased by a product-count instrument that no longer satisfies its exclusion restriction,
while the corrected likelihood remains close to the truth. We return to this in discussing our Medicare Advantage application. 

We also reemphasize at this point that recovering  $\sigma$ is not the same as recovering welfare, since congestion also enters mean utility. Under congestion, the likelihood correction alone keeps $\hat\sigma$ close to the truth, but a welfare calculation that omits the mean-utility congestion term overstates the gain from added
products.  Pairing the two corrections addresses both objects.   This is the
complementarity described in Section~\ref{sec:properties}: applications focused on substitution patterns and elasticities emphasize $\sigma$, but applications that  require welfare require estimates of $\tau$.  The  two corrections
address different objects, and a researcher who cares about welfare under congestion needs both.

\subsection{Robustness to non-normal errors}

Proposition~\ref{prop:gaussian} states that the corrected Gaussian likelihood is
exact under normal errors and a quasi-maximum-likelihood estimator otherwise,
consistent for the structural parameters when the quasi-score conditions hold. We
check the robustness of the point estimate directly. Holding the design at $\gamma = 0$ and $\sigma = 0.5$, with $J \sim U[1,10]$, we re-estimate the corrected likelihood when the structural error is drawn not from a normal but from a $t(5)$ and a $t(3)$ distribution, each scaled to the same variance.

\begin{table}[htbp]
\centering
\caption{QML Robustness of the Corrected Likelihood Estimator to Non-Normal Structural Errors}
\label{tab:sim_qml}
\begin{tabular}{lccc}
\toprule
Error distribution & Mean $\hat\sigma$ & Bias & RMSE \\
\midrule
Normal & 0.4999 & -0.0001 & 0.0058 \\
$t(5)$ & 0.4939 & -0.0061 & 0.0093 \\
$t(3)$ & 0.4847 & -0.0153 & 0.0193 \\
\bottomrule
\end{tabular}
\begin{minipage}{0.85\textwidth}
\footnotesize \vspace{4pt} \textit{Notes:} True $\sigma = 0.5$; single product nest, $J \sim U[1,10]$; $M = 1{,}000$ markets, $R = 200$ replications. The structural error is drawn from each distribution and rescaled to unit variance. Estimation maintains the Gaussian (corrected) likelihood throughout.
\end{minipage}
\end{table}

The point estimate is robust to the error distribution. Under normal and $t(5)$
errors it is effectively unbiased; under the heavier-tailed $t(3)$ a small
downward bias appears, of about $0.015$, with root mean squared error still under
$0.02$. The recovery of $\sigma$ does not rest on normality. Normality remains
relevant for inference rather than for the point estimate: as noted in
Section~\ref{sec:properties}, the curvature-based standard errors of the Gaussian
likelihood are unreliable under heteroskedasticity, and we report clustered and
bootstrap standard errors in the application accordingly.

%% ============================================================
%% APPLICATION 5.1 -- Setting and policy   [PASTE UNIT 1 of 4]
%% 5.2 (Data) is already in the document, immediately after this.
%% Full policy/behavioral discussion retained per author instruction
%% (to be cut later). Notation consistent with Sections 1, 3.
%% ============================================================

\section{Application: Medicare Advantage} \label{sec:application}

\subsection{Setting} \label{sec:app_setting}

Medicare Advantage allows Medicare beneficiaries to enroll in a privately
administered health plan in place of the public fee-for-service program  known as
Traditional Medicare. Under Medicare Advantage, the Centers for Medicare and
Medicaid Services (CMS) pays a private insurer a monthly capitated amount to
finance an enrollee's care, with payments set through a competitive bidding
process in which insurers submit bids for each county they serve
\citep{curto2021}. An insurer may offer several plans in the same county,
differing in premiums, cost-sharing, supplemental benefits, and provider
networks. By 2024 more than half of all Medicare beneficiaries were enrolled in a
Medicare Advantage plan, up from roughly a third a decade earlier.

For most of the program's history, CMS limited the degree to which an insurer
could offer similar plans in the same market. Under the ``meaningful difference''
requirement, CMS approved a plan's bid only if its benefit package was
substantially different from the insurer's other plans in the area along key
dimensions such as premium, cost-sharing, or benefits. The stated rationale was
grounded in consumer decision-making.  CMS argued that a large number of similar
options could overwhelm beneficiaries, discourage active choice, and entrench
incumbents, while  limiting near-duplicate plans would keep the menu manageable and reduce search costs. 

These concerns had support in the health economics literature, where several papers document systematic errors in Medicare plan choice,
including overweighting premiums relative to expected out-of-pocket costs
\citep{abaluck2011choice}, inertia and status-quo bias
\citep{handel2013adverse, ericson2014consumer}, and difficulty evaluating complex
insurance products \citep{bhargava2017choose}. More directly,
\citet{afendulis2015dominated} found that beneficiaries frequently failed to
switch into plans that strictly dominated their current coverage, and
\citet{mcwilliams2011complex} found that additional plan options raised enrollment
only when beneficiaries faced fewer than roughly fifteen plans.  Beyond that level, enrollment flattened or fell, especially among beneficiaries with impaired
cognition. The premise of the meaningful difference requirement was that, past
some point, more plans impaired choice.

In its Contract Year 2019 rulemaking, CMS changed its approach and  eliminated the
meaningful difference requirement \citep{cms2018rule}.
The agency characterized the requirement as an unnecessary limit on plan variety
that could push insurers to degrade benefits in order to keep their packages
formally distinct.  CMS expressed confidence that better decision-support tools
would help beneficiaries navigate a larger set of plans. The effect of the policy change on plan variety was
immediate. The number of plans available to a typical beneficiary roughly doubled
over the following six years. The growth came not
primarily from new insurers entering markets, but from existing insurers adding
plan variants within their current service areas.
%plans that differed marginally in supplemental benefits or cost-sharing, and sometimes more substantially in provider network, a dimension not visible on the Medicare Plan Finder.

Whether this proliferation raised or lowered consumer welfare is theoretically
ambiguous. In the   product-variety literature, additional varieties raise
welfare by matching consumers more closely to their preferences
\citep{dixit1977monopolistic}.  In differentiated-product markets the welfare gain
depends on whether new products expand the set of meaningfully distinct options or
 crowd the existing space, a distinction \citet{bew2016} formalize for radio
markets. Against the matching benefit, search costs can  soften competition if they make consumers less likely to
compare alternatives \citep{stigler1961economics, diamond1971model}, a channel of
particular concern for an elderly population facing complex insurance choices.  These frictions are large in practice.  \cite{YeoMiller2018}  estimate switching costs of \$1,600 to \$2,000 per enrollee in Medicare Part D from aggregate market share data, evidence that the costs of navigating plan menus are first-order in Medicare markets.
Which force dominates is an empirical question, and answering it requires a demand
model that correctly handles variation in the size of the choice set. 

The Medicare Advantage setting is also well suited to our corrected likelihood estimator. Plan counts vary widely across counties at a point in time and shift  within counties after the policy change. Variation across both place and time is what identifies the nesting and crowding parameters.

\subsection{Data} \label{sec:data}

\subsubsection*{Estimation Data}

The working data is a county--year panel of Medicare Advantage plans covering
2014--2024, the eleven years spanning the 2019 elimination of the meaningful
difference requirement. Each observation is a plan in a county in a year.  We record  the plan's enrollment, its premium and benefit characteristics, the county
benchmark payment rate, and the county's Medicare-eligible population. The panel
covers the fifty states and the District of Columbia and includes the plan types in the individual market (HMO, HMO point-of-service, local PPO, regional PPO,
and private fee-for-service) while excluding Special Needs Plans, employer group plans, and cost plans, none of which are offered to general beneficiaries on the same terms.\footnote{These exclusions are standard in demand studies of Medicare Advantage plan choice. See, for example, \citet{curto2021}. The data source for Medicare premiums covers only plans that offer Part~D drug coverage, so enrollees in Medicare Advantage plans with no Part~D coverage (3.6\% of MA enrollment) remain in the outside share with traditional Medicare. We also drop a few markets with missing eligibility counts due to county-code or reporting-region changes.}

A market is a county--year. Within a market, a beneficiary chooses among the
available plans or remains in traditional fee-for-service Medicare, which is the
outside option. We estimate the model under two nesting structures.  The first uses a single product nest containing all Medicare Advantage plans.  The single nest structure aligns with our theoretical structure and simulations, but imposes a strong assumption of symmetric substitutability among plan types.  For our primary specification, we group plans into two nests by plan type: an HMO
nest (HMO and HMO point-of-service) and a PPO nest (local PPO, regional PPO,
and private fee-for-service).  This structure groups plans that share a network and cost-sharing architecture.  The two-nest structure is the specification we carry through to welfare calculations.\footnote{Nested structures appear early in the health plan choice literature, with nests defined by institutional features: \citet{feldman1989} nest employer-sponsored plans by whether enrollees may freely choose their physician, \citet{atherly2004} nest the Medicare Advantage/fee-for-service decision above plan choice within the sector, and \citet{town2003} group Medicare HMOs by prescription drug coverage. We choose the HMO/PPO structure because it groups plans sharing a network and cost-sharing architecture, and because plan proliferation following the policy change was stronger in PPO plans.}

\begin{figure}[t]
    \centering
    \includegraphics[width=0.85\textwidth]{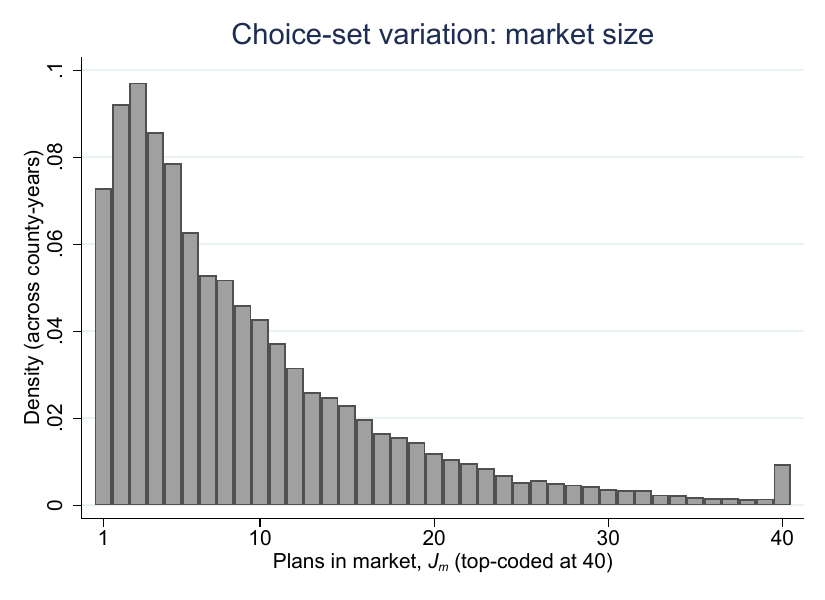}
    \caption{Choice-Set Variation Across Markets}
    \label{fig:ma_choiceset_market}
\end{figure}

\begin{figure}[t]
    \centering
    \includegraphics[width=0.85\textwidth]{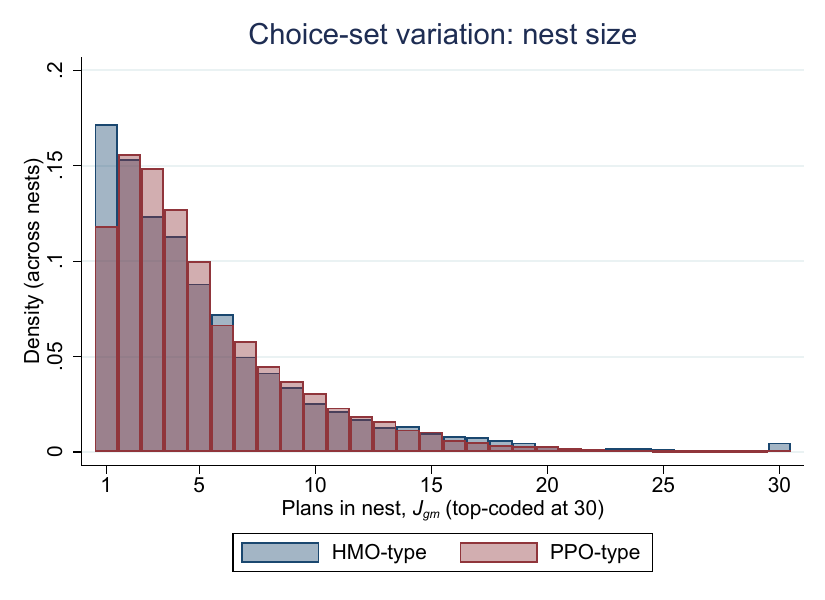}
    \caption{Choice-Set Variation by Plan Type}
    \label{fig:ma_choiceset_nest}
\end{figure}

The estimation sample contains 288{,}943 plan-county-year observations across  3{,}048 counties over the eleven years.\footnote{Elasticity and welfare calculations in Sections~\ref{sec:app_elasticities} and \ref{sec:app_welfare} use the estimation-eligible sample of 289{,}117 observations, which retains the 174 observations where the county benchmark is missing. The welfare counterfactual is then evaluated on the 2019--2024 markets with a valid 2018 plan count.}   The distribution of plans available across markets is illustrated in figure \ref{fig:ma_choiceset_market}. The number of plans in a market averages 9.2 and ranges from a single plan to
63, with a median of 7 and an interquartile range from 3 to 12.  The  distribution
is right-skewed, with a long tail of dense markets.  Most of this variation is
within rather than between plan types: the HMO and PPO nests each contain about five plans on average, and 71.7 percent of markets contain both
nests.  Markets with more than 40 plans are collected in the right tail in the figure.  The distribution of plan counts by type is illustrated in figure \ref{fig:ma_choiceset_nest}, with the right tail category containing markets with more than 30 plans of a particular type.

\begin{figure}[t]
    \centering
    \includegraphics[width=0.85\textwidth]{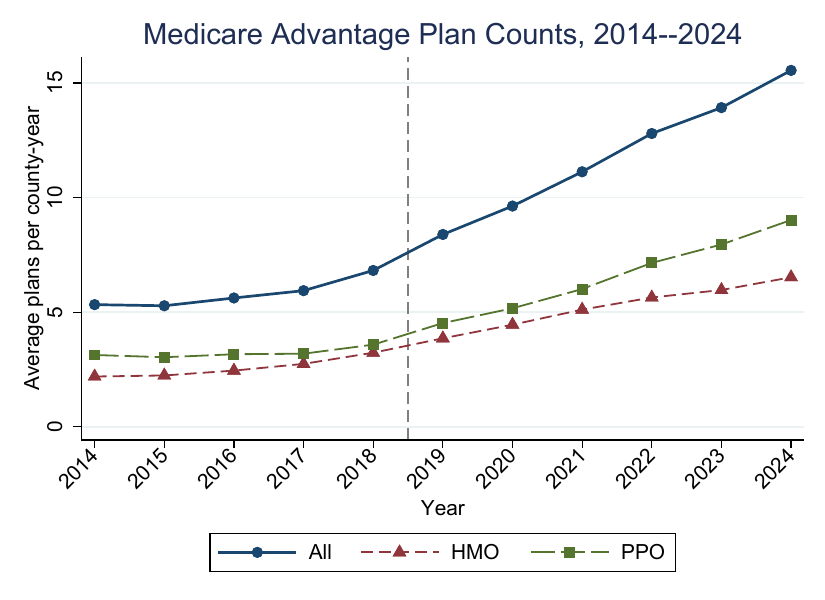}
    \caption{Plan Counts by Year, 2014--2024}
    \label{fig:ma_proliferation}
\end{figure}

The 2019 policy change is visible directly in the raw data, illustrated in figure \ref{fig:ma_proliferation}. After elimination of the meaningful difference requirement, the mean number of plans available in a market rose from
6.8 in 2018 to 15.5 in 2024, more than doubling over the six years following the change.

\subsubsection*{Data Construction}

The working data is assembled from four administrative sources, all published by
the Centers for Medicare and Medicaid Services and obtained either directly
or through the NBER archive.

\emph{Enrollment.} Plan enrollment by county comes from the CMS Monthly Enrollment
by Contract, Plan, State, and County (CPSC) files, using the December snapshot for
each year. Each release pairs a contract-information file which records the
organization type, plan type, and the special needs and employer-group indicators with an enrollment file recording county-level counts. We merge the
two on contract and plan identifiers. CMS suppresses enrollment counts of ten or fewer, which we treat as missing.

\emph{Premiums and benefits.} Plan premiums and benefit characteristics come from
the CMS Plan Premium Report, distributed with the Medicare Advantage and Part~D
Landscape Source files. These provide the Part~C monthly premium, the Part~D
premium components, the annual drug deductible, and indicators for supplemental
gap coverage, all at the plan--county--year level. They merge to the enrollment
panel on contract, plan, county, and year.

\emph{County benchmarks.} County benchmark payment rates come from the NBER
ratebook files, available for all years. The benchmark is the maximum monthly
per-beneficiary amount CMS will pay a plan before risk adjustment, set
administratively from historical fee-for-service spending in the county. We use the
base benchmark rate, excluding quality bonuses, as a supply-side instrument for the
plan premium.  The benchmark base rate shifts the plan's costs and entry incentives but is determined by
fee-for-service spending patterns rather than by the unobserved quality of any
individual plan.  Merging the ratebook to the panel requires a crosswalk from the
ratebook's Social Security Administration county codes to Federal Information
Processing Standards codes.

\emph{County penetration.} Total Medicare eligibles and aggregate Medicare
Advantage enrollment by county come from the CMS State/County Penetration files,
again using December of each year. The eligible count is the denominator for market
shares, and the difference between eligibles and total Medicare Advantage
enrollment defines the outside-good share.

From these sources we construct our estimation variables.   Plan $j$'s market
share is its enrollment divided by county Medicare eligibles, $s_{jm} =
\text{enrolled}_{jm}/\text{eligibles}_m$.  The outside-good share is $s_{0m} = 1 -
\sum_j s_{jm}$, the fraction of eligibles in traditional fee-for-service Medicare.
The within-nest share $s_{j|g}$ is plan $j$'s share among the plans in its own
nest. The dependent variable is the log-share ratio $y_{jm} = \ln s_{jm} - \ln
s_{0m}$. We drop markets in which the outside-good share is non-positive or equal to
one since for these markets the log-share ratio is undefined.

The price in our estimation is the Part~C monthly premium.  A notable feature of Medicare Advantage is that  a majority (60.6\%) of  plans charge no Part~C premium.  The mean premium is \$17.95 per month, but the median premium is zero.  A  growing share of plans also reduce the enrollee's Part~B premium through a giveback, a rebate financed out of the plan's bid.   Givebacks were offered by 2.6 percent of plan-years in 2018 and 13.2 percent by 2024, averaging \$64.80 per month among the plans that offer them.   

Because the Part~B premium lowers the effective price of enrollment, an alternative price measure for the estimation would be the premium less the  Part~B giveback.  We use the
premium rather than the net price because the net price cannot be
credibly instrumented.  The county benchmark is a strong instrument for the premium, but it is essentially uncorrelated with the net price once givebacks are netted
out.  This is ultimately because givebacks are funded from the same benchmark that instruments the premium.\footnote{The natural cost-based alternative, an instrument built from the
statutory rebate share, is strong in the first stage but fails the exclusion
restriction, since its power is derived from  the plan's star rating, which affects
demand directly.}  Because incorporating givebacks would lower the effective price
that enrollees face, the welfare magnitudes we report are
conservative, since accounting for givebacks would raise the surplus from any given choice set. 

\subsubsection*{Summary statistics}

Table~\ref{tab:summary_sample_structure} summarizes the data structure relevant to our single and dual nest structure.  Table~\ref{tab:summary_estimation_variables} reports summary statistics for  estimation variables.   The small plan market shares and the large outside-good share indicate the data are well suited to the  share inversion as the estimating object. The county benchmark is the instrument for the premium, and its cross-county variation is what identifies the price coefficient.  The wide dispersion in the number of plans per market is the variation that makes the likelihood correction vary across markets. Without it, the Jacobian contribution would be common across markets, but it would still depend on $\sigma$ and therefore still affect the likelihood estimate. With variable choice sets, the correction also changes how markets with different numbers of plans contribute to the likelihood and connects directly to the product-count variation used by the standard count-IV estimator.

% T_summary_sample_structure.tex  -- auto-generated by S10_summary_stats_tables.do
% READ-ONLY frozen sample; do not hand-edit numbers.
\begin{table}[htbp]\centering
\caption{Sample Structure}
\label{tab:summary_sample_structure}
\begin{tabular}{lcccc}
\toprule
 & All MA & HMO & PPO & HMO/PPO Nests \\
\midrule
N Plans (Plan-County-Year) &   288,943 &   127,815 &   161,128 &   288,943 \\
N Markets (County-Year) &    31,281 &    23,437 &    30,268 &    31,281 \\
N Nests &    31,281 &    23,437 &    30,268 &    53,705 \\
Mean Plans per Market/Nest &    9.24 &    5.45 &    5.32 &    5.38 \\
Mean Log Plans per Market/Nest &   1.844 &   1.332 &   1.384 &   1.361 \\
\bottomrule
\end{tabular}
\end{table}

% T_summary_estimation_variables.tex  -- auto-generated by replication/estimation/01b_summary_stats.do
% READ-ONLY frozen sample; do not hand-edit numbers. Two nest definitions, same sample.
\begin{table}[htbp]\centering
\caption{Summary Statistics for Estimation Variables}
\label{tab:summary_estimation_variables}
\begin{tabular}{lcc}
\toprule
 & All MA Nest & HMO/PPO Nests \\
\midrule
Dependent variable, $\ln s_j-\ln s_0$ &  -4.575 &  -4.575 \\
 & ( 1.391) & ( 1.391) \\
Plan share, $s_j$ &  0.0180 &  0.0180 \\
 & (0.0259) & (0.0259) \\
Outside share, $s_0$ &  0.7719 &  0.7719 \\
 & (0.1060) & (0.1060) \\
Within-nest share, $s_{j\mid g,m}$ &  0.1083 &  0.1859 \\
 & (0.1622) & (0.2333) \\
Part C premium, $p_j$ (\$/mo) &   17.95 &   17.95 \\
 & ( 33.33) & ( 33.33) \\
Drug deductible (\$) &  140.02 &  140.02 \\
 & (161.16) & (161.16) \\
Gap coverage &   0.493 &   0.493 \\
 & ( 0.500) & ( 0.500) \\
HMO indicator &   0.442 &   0.442 \\
 & ( 0.497) & ( 0.497) \\
County benchmark (\$/mo) &  961.44 &  961.44 \\
 & (125.97) & (125.97) \\
Plan count & $J_m$:   16.61 & $J_{gm}$:    9.42 \\
 & ( 11.26) & (  6.97) \\
\bottomrule
\end{tabular}
\par\smallskip
\footnotesize\begin{minipage}{\textwidth}
\emph{Notes.} Means with standard deviations in parentheses, over the estimation
sample. The two columns describe the same observations under the two nest
definitions: a single Medicare Advantage nest, and separate HMO and PPO nests.
In the within-nest share row, $g$ denotes the single Medicare Advantage nest
under the single-nest definition and the plan's HMO or PPO nest under the
two-nest definition. The plan-count row reports the market total $J_m$ under
the single-nest definition and the nest-level count $J_{gm}$ under the two-nest
definition. Both are averaged across plan observations, so larger markets
contribute more rows; the market-level averages reported in
Table~\ref{tab:summary_sample_structure} are correspondingly smaller.
\end{minipage}
\end{table}

\FloatBarrier

\subsection{Estimation and results} \label{sec:app_estimates}

We estimate the nested logit of plan choice by two methods on the same data, the
same fixed effects, and the same price instrument, so that differences between estimators reflect how each treats the within-nest share. Our baseline model includes a single product nest for consistency with our modeling and simulation framework.  This framework imposes symmetric substitution across all Medicare Advantage plans conditional on observed characteristics.  Including an HMO indicator allows differences in plan types to enter mean utility, but the single nest approach ultimately asks a single parameter $\sigma$ to capture the closeness of all Medicare Advantage plans relative to traditional Medicare.   The  baseline estimating
equation  is
\begin{equation} \label{eq:app_est}
\ln s_{jm} - \ln s_{0m} = X_{jm}\beta + \alpha\, p_{jm}
+ \sigma \ln s_{j|g} + \xi_{jm},
\end{equation}
where $p_{jm}$ is the Part~C premium, $X_{jm}$ are plan characteristics (the
drug deductible, HMO/PPO indicator and a gap-coverage indicator), and $\sigma$ is the substitution  parameter of primary interest.   We include county fixed effects and
year dummies and instrument the premium with the county benchmark rate. 

We compare the standard linear estimator (labeled count IV throughout)  to the corrected likelihood estimator.  The
count IV estimator estimates (\ref{eq:app_est}) by two-stage least
squares, instrumenting the within-nest share with the count of rival
plans in the nest. The corrected likelihood estimator maximizes the likelihood that adds the Jacobian
term $\sum_m\sum_g (J_{gm}-1)\ln(1-\sigma)$, and uses no instrument for the
within-nest share.

We note that for the count IV estimator we include only a single instrument for the within-nest share, the count of rival plans in the nest.  We recognize that many applications with count instruments include more than simple counts of rival products, for example combining in-market counts with out-of-market counts or averages.  We adopt the single instrument to maintain comparability with our theoretical treatment and simulations, and to highlight identification issues that arise with count instruments. To the extent that additional instruments draw on variation independent of in-nest product counts, our estimates will overstate the difference between the count IV and corrected likelihood estimator.\footnote{In this application the leading alternative, a leave-out count of plans offered by the same parent organization in other counties, is strongly correlated with the in-market count within each nest. It is therefore not a source of variation independent of in-market plan counts, though it is not a pure proxy for them either.}

\subsubsection*{Combined MA Estimates (Single Nest)}

Table~\ref{tab:single_nest_params_paper} reports results of the baseline single-nest model.  The first two columns report results with no congestion parameter, while the third column includes the crowding term.   The first point of note is that with a single nest, count IV produces an invalid nesting parameter of  $\hat\sigma=1.23$.  The excluded instrument strongly predicts the within-nest share, but the resulting IV estimate is outside of the  admissible nested-logit interval $(0,1)$.  

% T8_single_nest_params_PAPER.tex -- GENERATED by T7_tab5_tab8_generators.do,
% 2026-07-29. Supersedes the hand-typeset version.
% Point estimates: S8_single_nest_estimates.txt blocks (4),(5),(6).
% Control coefficients: S8x_control_extract.txt blocks (1),(2),(3).
% sigma SEs cols 2-3: county block bootstrap B=200 (single-nest summary).
% tau SE col 3: trimmed county block bootstrap B=200 (see report).
\begin{table}[htbp]\centering
\caption{Single-Nest Medicare Advantage Model}
\label{tab:single_nest_params_paper}
\begin{tabular}{lccc}
\toprule
 & Count IV & Corrected & Corrected Likelihood \\
 & (2SLS) & Likelihood & + Crowding \\
\midrule
$\alpha$ (premium) & $-0.083$ & $-0.058$ & $-0.072$ \\
 & $(0.025)$ & $(0.009)$ & $(0.011)$ \\
$\sigma$ (nesting) & $1.230$\textsuperscript{\dag} & $0.528$ & $0.508$ \\
 & $(0.116)$ & $(0.006)$ & $(0.007)$ \\
$\tau$ (crowding, $\ln J_m$) & --- & --- & $-0.582$ \\
 &  &  & $(1.004)$\textsuperscript{\ddag} \\
Drug deductible & $-0.0012$ & $-0.0016$ & $-0.0019$ \\
 & $(0.0004)$ & $(0.0002)$ & $(0.0002)$ \\
Gap coverage & $-0.135$ & $-0.037$ & $-0.054$ \\
 & $(0.044)$ & $(0.013)$ & $(0.015)$ \\
HMO indicator & $-0.944$ & $-0.559$ & $-0.707$ \\
 & $(0.288)$ & $(0.098)$ & $(0.119)$ \\
\midrule
County, Year Fixed Effects & Yes & Yes & Yes \\
Observations & $288{,}943$ & $288{,}943$ & $288{,}943$ \\
Markets & $31{,}281$ & $31{,}281$ & $31{,}281$ \\
\bottomrule
\end{tabular}
\par\smallskip
\footnotesize\begin{minipage}{\textwidth}
\emph{Notes.} Medicare Advantage demand with a single inside nest and Traditional
Medicare as the outside option. All columns include county and year fixed effects.
Standard errors for corrected likelihood estimates of $\sigma$
 and $\tau$ are county block bootstrap standard errors ($B=200$).
\textsuperscript{\dag}~Inadmissible: outside $[0,1)$.
\textsuperscript{\ddag}~95 percent bootstrap percentile interval
$[-1.12, -0.39]$. See text for details.
\end{minipage}
\end{table}

With the corrected likelihood and no share instrument, the substitution parameter is estimated precisely at $\sigma=0.528$. Adding a parametric control for crowding has minimal impact on the substitution parameter: estimates of $\sigma$ are essentially unchanged at $\sigma=0.508$.  

The crowding parameter in the third column is negative but imprecisely estimated, a further indication that the single-nest structure is the wrong structure for welfare calculations. The price coefficient is negative and precisely estimated through all specifications.

We note that substitution parameters are estimated more precisely under the corrected likelihood than count IV and return to this topic in the two-nest case. 

\subsubsection*{HMO/PPO Estimates (Two Product Nests)}

The count IV estimate of  $\sigma>1$ in the single nest case precludes welfare calculations of the policy change and limits comparisons across the two estimators, suggesting that a  richer substitution
structure may be more informative.  We extend the likelihood correction to the case of separate product nests for PPO and HMO Medicare Advantage plans.  

With multiple product nests, the log-Jacobian of the share transformation is additively separable across nests, so the correction is applied nest by nest and summed,
\begin{equation}
  \sum_m \sum_g \bigl(J_{gm}-1\bigr)\ln\!\bigl(1-\sigma_g\bigr),
  \label{eq:jacobian_twonest}
\end{equation}
where $J_{gm}$ is the number of plans in nest $g$ in market $m$ and $\sigma_g$ is the nest-specific substitution parameter. In our Medicare Advantage application, $g\in\{\mathrm{HMO},\mathrm{PPO}\}$.  The corresponding estimating equation is
\begin{equation}
y_{jgm}
=
x_{jgm}\beta
+
\alpha p_{jgm}
+
\sigma_g \ln s_{j\mid g,m}
+
\tau r(J_{gm})
+
\xi_{jgm},
\label{eq:twonest_tau}
\end{equation}
with $r(J_{gm})=\ln J_{gm}$ as in the baseline specification. We maintain a common crowding parameter $\tau$ identified from variation in HMO and PPO nest sizes conditional on the other controls and fixed effects. With county and year fixed effects, the identifying variation is within-county movement in nest sizes over time and across plan types, concentrated in the post-2019 expansion. We specify a single crowding parameter common to both nests: each nest's plan count crowds its own product space, and the common $\tau$ imposes the same per-plan penalty across types.\footnote{The restriction is readily extended to nest-specific crowding parameters, although in applications with fixed effects it does place demands on the data, requiring sufficient variation across types as well as across markets.} The corrected likelihood allows identification of $\sigma_g$ through the nest-specific Jacobian terms in \eqref{eq:jacobian_twonest}, while the product-count terms enter mean utility directly through $\tau r(J_{gm})$.

\begin{table}[htbp]\centering
\caption{Two-Nest Medicare Advantage Model with Common Crowding}
\label{tab:two_nest_common_tau_paper}
\begin{tabular}{lccc}
\toprule
 & Count IV & Corrected & Corrected Likelihood \\
 & (2SLS) & Likelihood & + Common Crowding \\
\midrule
$\alpha$ (premium) & $-0.033$ & $-0.034$ & $-0.034$ \\
 & $(0.010)$ & $(0.008)$ & $(0.008)$ \\
$\sigma_{\mathrm{HMO}}$ & $0.323$ & $0.140$ & $0.136$ \\
 & $(0.066)$ & $(0.008)$ & $(0.009)$ \\
$\sigma_{\mathrm{PPO}}$ & $0.294$ & $0.122$ & $0.117$ \\
 & $(0.041)$ & $(0.008)$ & $(0.008)$ \\
$\tau$ (crowding, $\ln J_{gm}$) & --- & --- & $-0.052$ \\
 &  &  & $(0.026)$ \\
Drug deductible & $-0.0013$ & $-0.0015$ & $-0.0015$ \\
 & $(0.0001)$ & $(0.0001)$ & $(0.0001)$ \\
Gap coverage & $0.012$ & $0.032$ & $0.031$ \\
 & $(0.020)$ & $(0.013)$ & $(0.013)$ \\
\midrule
County, Year Fixed Effects & Yes & Yes & Yes \\
Observations & $288{,}943$ & $288{,}943$ & $288{,}943$ \\
Markets & $31{,}281$ & $31{,}281$ & $31{,}281$ \\
\bottomrule
\end{tabular}
\par\smallskip
\footnotesize\begin{minipage}{\textwidth}
\emph{Notes.} Medicare Advantage demand model with HMO and PPO nests,
Traditional Medicare as the outside option.  All specifications include county and
year fixed effects. Standard errors for corrected likelihood estimates of $\sigma$ and $\tau$
are county block bootstrap standard errors ($B=200$).
See text for details.
\end{minipage}
\end{table}

Table \ref{tab:two_nest_common_tau_paper} reports parameter estimates with two product nests.  The count IV specification produces valid substitution parameters with two product nests, allowing welfare comparison across estimators.  Count IV estimates of substitution parameters are more than double the magnitude of the parameters estimated with corrected likelihood, implying that 
 plans of the same type substitute less
readily for one another than the count IV estimator detects. The  corrected estimates have bootstrap standard errors of $0.008$ to $0.009$.   A gap of this size  is
well outside of sampling variation and consistent with congestion of product space as illustrated in the simulations.\footnote{As noted in
Section~\ref{sec:properties}, the Gaussian curvature understates uncertainty when the errors are not homoskedastic normal.  We report instead the county block-bootstrap standard error,  which accounts for the within-county dependence of the
data. Even at this more conservative figure, the difference between the corrected likelihood and count IV estimates of $\sigma$ is large.}

The price coefficient is negative in all three columns and is similar in magnitude at $-0.033$.  This is expected since  $\alpha$ is identified from the price instrument and the likelihood correction does not impact that channel. The estimators differ on the
nesting parameter. 

A likelihood-ratio test confirms the nesting structure is not an innocuous
choice. The restriction $\sigma_{HMO}=\sigma_{PPO}=0$ collapses the nested
logit to plain logit; the statistic is 2,147 against the conservative
Kodde--Palm five percent bound of 5.14 \citep{koddepalm1986}. The data
reject the plain logit.
%\FloatBarrier

\subsection{Elasticities} \label{sec:app_elasticities}

The substitution parameters have a direct behavioral interpretation. The
estimate $\sigma_{HMO}=0.136$ means that when an HMO plan loses an enrollee,
the enrollee is more likely to move to another HMO than the plan's overall
market share alone would predict, but only moderately so; most of the
correlation in preferences among same-type plans is captured by observed
characteristics and the county and year fixed effects. The PPO estimate,
$\sigma_{PPO}=0.117$, implies slightly weaker within-type clustering. Values
this far from one mean that Medicare Advantage plans of the same type are
meaningfully differentiated: an additional HMO plan in a county is not simply
splitting a fixed HMO clientele. The count IV estimates, $\sigma_{HMO}=0.323$
and $\sigma_{PPO}=0.294$, more than double both parameters, describing a market
in which same-type plans are much closer substitutes and new plans mostly
cannibalize their own type. The two estimators thus disagree about the basic
structure of competition, and the simulations attribute the disagreement to
product-count variation entering demand directly.

% tab5_elasticities.tex -- GENERATED by T7_tab5_tab8_generators.do, 2026-07-29.
% Supersedes the hand-typeset version. Elasticities evaluated AT nest-specific
% sample means (not means of plan-level elasticities). Formulas:
%   own    =  (alpha/(1-s_g))*(1 - s_g*s_jg - (1-s_g)*s_j)*p     (signed negative)
%   within = -(alpha/(1-s_g))*(s_g*s_jg + (1-s_g)*s_j)*p
%   cross  = -alpha*s_j*p
% The outside-good row was REMOVED 2026-07-29: it had been computed as
% -alpha*s_0*p, which is not the elasticity of any share. The correct object,
% d ln s_0 / d ln p_j = -alpha*s_j*p, equals the cross-nest row exactly.
\begin{table}[htbp]\centering
\caption{Demand Elasticities at Sample Means by Estimator and Plan Type}
\label{tab:elasticities}
\begin{tabular}{lcccc}
\toprule
 & \multicolumn{2}{c}{Count IV} & \multicolumn{2}{c}{Corrected Likelihood} \\
\cmidrule(lr){2-3}\cmidrule(lr){4-5}
 & HMO & PPO & HMO & PPO \\
\midrule
Own-price elasticity & $-0.613$ & $-0.955$ & $-0.497$ & $-0.791$ \\
Within-nest cross-price & $0.048$ & $0.069$ & $0.022$ & $0.030$ \\
Cross-nest cross-price & $0.009$ & $0.012$ & $0.009$ & $0.012$ \\
\midrule
Mean within-nest share $s_{j\mid g}$ & $0.183$ & $0.188$ & $0.183$ & $0.188$ \\
Mean plan share $s_j$ & $0.020$ & $0.017$ & $0.020$ & $0.017$ \\
Mean premium (dollars per month) & $13.35$ & $21.59$ & $13.35$ & $21.59$ \\
Observations & $127{,}908$ & $161{,}209$ & $127{,}908$ & $161{,}209$ \\
\bottomrule
\end{tabular}
\par\smallskip
\footnotesize\begin{minipage}{\textwidth}
\emph{Notes.} Elasticities evaluated at the nest-specific sample means
reported in the lower panel, using the estimates in
Table~\ref{tab:two_nest_common_tau_paper}. Columns index the plan whose
premium changes; rows index the responding alternative. Own-price
elasticities are signed negative, substitution terms positive. See text for
details.
\end{minipage}
\end{table}

Table~\ref{tab:elasticities} translates the nesting parameters into elasticities at
sample means. Own-price elasticities are small under both estimators, as expected in a
market where 60.6 percent of plans charge no Part C premium and price
competition operates through benefits and rebates.\footnote{Elasticities are
evaluated on the full estimation-eligible sample, which slightly exceeds the
estimation sample because it does not require the price instrument to be
non-missing.} The corrected likelihood estimates are
$-0.497$ for HMO plans and $-0.791$ for PPO plans; the count IV estimates are
larger in magnitude, $-0.613$ and $-0.955$, because a larger $\sigma_g$ scales up
the within-nest response to a plan's own price. The largest difference in estimators appears
in the within-nest cross-price row. Under the corrected likelihood, a same-type rival's
price increase raises a plan's demand by roughly two and a half times as much
as a cross-type rival's.  Under count IV the within-type responses are inflated in
proportion to the larger $\sigma_g$, so the estimator overstates how much of a
plan's enrollment is recaptured within its own type and understates diversion
across types and to the outside option. These are the substitution patterns
most relevant to merger review, so inflation using count IV is not innocuous even before any welfare calculation.

\FloatBarrier
% ------------------------------------------------------------

\subsection{Welfare} \label{sec:app_welfare}

The policy question is whether the plan proliferation that followed the
elimination of the meaningful difference requirement raised consumer welfare.
We answer it with a counterfactual that restores each county's 2018 plan count
by removing the lowest-enrollment plans and computes the change in consumer
surplus between the observed and counterfactual menus. The analysis covers the
16,562 county-year markets from 2019 through 2024 with a valid 2018 plan count.

Consumer surplus in a nested logit market is
\begin{equation} \label{eq:app_cs}
CS_m = \frac{1}{-\alpha}\,
\ln\!\left[\,1 + \sum_{g}\Bigl(\sum_{j \in g}
e^{\,(v_{jgm} + \tau \ln J_{gm})/(1-\sigma_g)}\Bigr)^{1-\sigma_g}\right],
\end{equation}
where $v_{jgm}$ is mean utility excluding the product-count term and $\tau \ln
J_{gm}$ enters mean utility directly. Three parameters affect the calculation. The
price coefficient $\alpha$ converts utility to dollars. The substitution
parameters $\sigma_g$ govern how much of an added plan's enrollment reflects business
stealing inside the nest. The crowding parameter $\tau$ discounts mean utility
as plan counts expand, so removing plans in the counterfactual raises the utility of
the plans that remain. All parameters are held at their estimated values.

% tab7_welfare.tex -- GENERATED by T6_welfare_tables.do, 2026-07-29.
% Supersedes the hand-typeset version. Three columns.
% Engine compute_cs2; variant-A counterfactual (restore each county's 2018 plan
% count by removing lowest-enrollment plans). Delta CS per Medicare ELIGIBLE/month.
%   col1 Count IV        sigma 0.3234/0.2935 alpha -0.0335 tau 0
%   col2 Corrected       sigma 0.140 /0.122  alpha -0.0337 tau 0
%   col3 Corrected+crowd sigma 0.136 /0.117  alpha -0.0336 tau -0.052
\begin{table}[htbp]\centering
\caption{Welfare Value of Post-2019 Medicare Advantage Plan Growth by Estimator}
\label{tab:welfare_preferred}
\begin{tabular}{lccc}
\toprule
 & Count IV & Corrected & Corrected Likelihood \\
 & (2SLS) & Likelihood & + Crowding \\
 & ($\tau{=}0$) & ($\tau{=}0$) & ($\tau{=}{-}0.052$) \\
\midrule
\multicolumn{4}{l}{\emph{Panel A: 2024}} \\
\quad Mean $\Delta CS$, eligible-weighted   & $0.795$ & $0.952$ & $0.665$ \\
\quad Mean $\Delta CS$, enrollment-weighted & $0.780$ & $0.939$ & $0.619$ \\
\quad Mean $\Delta CS$, unweighted          & $1.225$ & $1.434$ & $1.193$ \\
\quad Share of markets with $\Delta CS>0$   & $0.969$ & $0.969$ & $0.948$ \\
\quad Share of markets with $\Delta CS<0$   & $0.000$ & $0.000$ & $0.020$ \\
\quad Aggregate 2024 (millions of dollars per year) & $620$ & $742$ & $519$ \\
\quad Markets & $2{,}757$ & $2{,}757$ & $2{,}757$ \\
\addlinespace
\multicolumn{4}{l}{\emph{Panel B: pooled 2019--2024}} \\
\quad Mean $\Delta CS$, eligible-weighted   & $0.443$ & $0.533$ & $0.342$ \\
\quad Mean $\Delta CS$, enrollment-weighted & $0.449$ & $0.543$ & $0.320$ \\
\quad Mean $\Delta CS$, unweighted          & $0.678$ & $0.800$ & $0.650$ \\
\quad Share of markets with $\Delta CS>0$   & $0.876$ & $0.876$ & $0.824$ \\
\quad Share of markets with $\Delta CS<0$   & $0.000$ & $0.000$ & $0.052$ \\
\quad Markets & $16{,}562$ & $16{,}562$ & $16{,}562$ \\
\bottomrule
\end{tabular}
\par\smallskip
\footnotesize\begin{minipage}{\textwidth}
\emph{Notes.} Consumer surplus in dollars per Medicare eligible per month.
Each market's menu is reduced to its 2018 plan count by removing the
lowest-enrollment plans. All three columns use one common surplus
engine, the same counterfactual, and the same markets, and differ only
in the demand parameters, taken from Table~\ref{tab:two_nest_common_tau_paper}.
See text for details.
\end{minipage}
\end{table}

Table~\ref{tab:welfare_preferred} reports the counterfactual under three
specifications that isolate the two sources of error. Column (1) uses
the two-nest count-IV estimates and sets $\tau=0$.  Column (2) uses the corrected-likelihood estimates and sets $\tau=0$. Column 3 uses the corrected-likelihood estimates with the estimated value of $\tau$.

The comparison of the second and third columns shows the cost of ignoring
crowding when substitution is estimated by the corrected likelihood. With $\tau=0$, the model
attributes \$0.95 per Medicare eligible per month to proliferation in 2024;
with the estimated crowding term, \$0.67.  Assuming away crowding overstates
the welfare gain by 43 percent. The overstatement is 20 percent unweighted.
Weighting matters because the markets that added the most plans are also the
largest, and these are the markets where the crowding discount is greatest
relative to the variety gain.  For scale, the preferred estimate is about four percent of the mean Part~C
premium, or roughly eight dollars per eligible per year.  Aggregated over the
sample counties it is the \$518.5 million figure reported below.

A model without crowding cannot represent a welfare loss from added variety.
With $\tau=0$ the counterfactual menu is a strict subset of the observed
menu, so consumer surplus cannot fall and no market can be made worse off.
Surplus falls in a market only when the crowding term is present.  In the preferred specification, surplus falls in 2.0\% of markets.

The first column of Table~\ref{tab:welfare_preferred} shows that the count
IV error is not simply a larger version of the same bias. The count-IV estimate lies between the two corrected-likelihood estimates because two biases operate in opposite directions in this application. Its larger $\sigma_g$ reduces the estimated gain from added variety, while imposing $\tau=0$ increases it. The partial offset is specific to these estimates; count IV still implies nonnegative welfare effects in every market.

The direction of the crowding correction is stable; its magnitude is not.
Table~\ref{tab:welfare_bootstrap} reports percentile intervals from the county
block bootstrap behind the two-nest estimates, recomputing the counterfactual
at each replicate's parameters. The crowding parameter is negative in 99.5
percent of replications, and the correction reduces the estimated welfare gain
in 98.5 percent. The dollar magnitudes are wider: the 2024 eligible-weighted
mean ranges from \$0.27 to \$0.88 per eligible per month, and the share of
losing markets from zero to 15 percent. Dollar figures are conditional on the
estimated price coefficient, which scales surplus but leaves the overstatement
ratio and the sign shares unchanged. Table~\ref{tab:tau_sensitivity} isolates
the crowding channel, varying $\tau$ alone at the preferred substitution
parameters: one bootstrap standard deviation moves the overstatement from 18
to 87 percent.

% tab8_welfare_bootstrap.tex -- generated by T5b_write_tab8.do, 2026-07-29.
% Source: T5_welfare_bootstrap.do, which reuses the B=200 county block bootstrap
% draws from the Table 7 estimation (S6_bootstrap_two_sigma_reps.dta; bsample
% cluster(fipscode), seed 20260611). No new blocks drawn, no re-estimation.
% Counterfactual: variant A (lowest-enrollment removal, market-level 2018 target).
% Delta CS is per Medicare ELIGIBLE per month.
\begin{table}[htbp]\centering
\caption{Bootstrap Uncertainty in the Welfare Counterfactual}
\label{tab:welfare_bootstrap}
\begin{tabular}{lccc}
\toprule
 & Estimate & \multicolumn{2}{c}{95\% percentile interval} \\
\cmidrule(lr){3-4}
 &  & 2.5\% & 97.5\% \\
\midrule
\multicolumn{4}{l}{\emph{Panel A: 2024}} \\
\quad Mean $\Delta CS$, eligible-weighted   &  0.665 &  0.270 &  0.882 \\
\quad Mean $\Delta CS$, enrollment-weighted &  0.619 &  0.181 &  0.860 \\
\quad Mean $\Delta CS$, unweighted          &  1.193 &  0.849 &  1.377 \\
\quad Share of markets with $\Delta CS>0$   &  0.949 &  0.818 &  0.968 \\
\quad Share of markets with $\Delta CS<0$   &  0.020 &  0.000 &  0.151 \\
\quad Crowding overstatement ratio                  &  1.442 &  1.054 &  2.764 \\
\addlinespace
\multicolumn{4}{l}{\emph{Panel B: pooled 2019--2024}} \\
\quad Mean $\Delta CS$, eligible-weighted   &  0.342 &  0.082 &  0.486 \\
\quad Mean $\Delta CS$, enrollment-weighted &  0.320 &  0.017 &  0.488 \\
\quad Mean $\Delta CS$, unweighted          &  0.650 &  0.439 &  0.764 \\
\quad Share of markets with $\Delta CS>0$   &  0.824 &  0.656 &  0.873 \\
\quad Share of markets with $\Delta CS<0$   &  0.052 &  0.004 &  0.220 \\
\quad Crowding overstatement ratio                  &  1.569 &  1.015 &  3.246 \\
\bottomrule
\end{tabular}
\par\smallskip
\footnotesize\begin{minipage}{\textwidth}
\emph{Notes.} Percentile intervals from the 200-replicate county block
bootstrap behind Table~\ref{tab:two_nest_common_tau_paper}, with the welfare
counterfactual recomputed at each replicate's parameters. $\Delta CS$ is in
dollars per Medicare eligible per month. The overstatement ratio is the mean
$\Delta CS$ at $\tau{=}0$ divided by the mean at the replicate's $\tau$,
holding $\sigma_g$ fixed. The price coefficient is held fixed; it scales
dollar levels but not the ratio or the sign shares.
\end{minipage}
\end{table}

% tab9_tau_sensitivity.tex -- GENERATED by T6_welfare_tables.do, 2026-07-29.
% tau in {0, -0.026, -0.052, -0.078} = point estimate +/- one bootstrap SD and zero.
% sigma held at the corrected values 0.136/0.117, alpha -0.0336. 2024, per eligible,
% eligible-weighted. Bootstrap interval from the reused B=200 draws (T5).
\begin{table}[htbp]\centering
\caption{Sensitivity of the Welfare Counterfactual to the Crowding Parameter}
\label{tab:tau_sensitivity}
\begin{tabular}{lcccc}
\toprule
 & $\tau{=}0$ & $\tau{=}{-}0.026$ & $\tau{=}{-}0.052$ & $\tau{=}{-}0.078$ \\
 &  & $(\hat{\tau}{+}s)$ & $(\hat{\tau})$ & $(\hat{\tau}{-}s)$ \\
\midrule
Mean $\Delta CS$ per eligible per month & $0.959$ & $0.813$ & $0.665$ & $0.514$ \\
\quad 95\% bootstrap interval at $\hat{\tau}$ &  &  & $[0.270,0.882]$ &  \\
Share of markets with $\Delta CS<0$ & $0.000$ & $0.003$ & $0.020$ & $0.063$ \\
\quad 95\% bootstrap interval at $\hat{\tau}$ &  &  & $[0.000,0.151]$ &  \\
Ratio to $\tau{=}0$ & $1.000$ & $1.179$ & $1.442$ & $1.865$ \\
\quad 95\% bootstrap interval at $\hat{\tau}$ &  &  & $[1.054,2.764]$ &  \\
\bottomrule
\end{tabular}
\par\smallskip
\footnotesize\begin{minipage}{\textwidth}
\emph{Notes.} Substitution parameters held at the preferred estimates;
columns vary the crowding parameter alone. Figures are for 2024, in dollars
per Medicare eligible per month, weighted by eligibles. The ratio row divides
the $\tau{=}0$ mean by the mean at each $\tau$. Bracketed intervals are 95
percent percentile intervals from the county block bootstrap, which varies
$\sigma_g$ and $\tau$ jointly.
\end{minipage}
\end{table}

Table~\ref{tab:welfare_quintiles} breaks the 2024 counterfactual out by
quintile of the market plan count. Gains under the preferred specification
rise from \$0.89 per eligible per month in the thinnest quintile to \$1.13 in
the second, then fall to \$0.44 in the densest, where the crowding discount is
largest. The overstatement from ignoring crowding rises monotonically with
plan counts, from 13 percent in the thinnest quintile to 72 percent in the
densest. The densest quintile contains 30 of the 56 losing markets and 62 percent of sample eligibles; it accounts for most of the aggregate overstatement and losses.

% tab10_welfare_quintiles.tex -- GENERATED by T6c_welfare_quintiles.do, 2026-07-29.
% Engine: compute_cs2, cloned verbatim from T6_welfare_tables.do (L60-82).
% Counterfactual: variant A -- each market's menu reduced to its 2018 plan count
% by removing the lowest-enrollment plans (T6_welfare_tables.do L117-121).
% Sample: the 2024 markets of the Table 9 Panel A counterfactual set.
% Parameters, locked, from T6_welfare_tables.do L54-57 (preferred column of
% tab7_welfare.tex; sources S7b_pointcheck / S9c_two_sigma_countiv):
%   sigma_HMO 0.136  sigma_PPO 0.117  alpha -0.0336
%   rows vary tau alone: -0.052 (preferred) and 0 (no crowding).
% Pure-channel convention of tab9_tau_sensitivity.tex, NOT Table 9 cols 2-3.
% Quintiles: xtile nq(5) on the 2024 market plan count; integer ties give
% uneven cells. Delta CS per Medicare ELIGIBLE per month, eligible-weighted.
% Reconciled to T6_welfare_tables.txt before writing: 0.6652 / 0.9590 / 0.0203.
\begin{table}[htbp]\centering
\caption{Welfare Effects by Market Plan-Count Quintile}
\label{tab:welfare_quintiles}
\begin{tabular}{lccccc}
\toprule
 & Q1 & Q2 & Q3 & Q4 & Q5 \\
  & $\bar{J}{=}5.1$ & $\bar{J}{=}10.6$ & $\bar{J}{=}14.9$ & $\bar{J}{=}20.6$ & $\bar{J}{=}33.8$ \\
\midrule
Mean $\Delta CS$, $\tau{=}{-}0.052$ (preferred) & $0.887$ & $1.128$ & $1.060$ & $1.014$ & $0.436$ \\
Mean $\Delta CS$, $\tau{=}0$ & $1.002$ & $1.350$ & $1.299$ & $1.331$ & $0.751$ \\
Ratio to $\tau{=}0$ & $1.130$ & $1.198$ & $1.225$ & $1.312$ & $1.721$ \\
\addlinespace
Share of markets with $\Delta CS<0$ (preferred) & $0.010$ & $0.004$ & $0.010$ & $0.022$ & $0.059$ \\
Share of markets with unchanged menus & $0.147$ & $0.000$ & $0.000$ & $0.000$ & $0.000$ \\
\addlinespace
Markets & $586$ & $524$ & $581$ & $556$ & $510$ \\
\bottomrule
\end{tabular}
\par\smallskip
\footnotesize\begin{minipage}{\textwidth}
\emph{Notes.} Quintiles of the 2024 market plan count; column headers give
the mean plan count in the quintile. Integer plan counts tie across quintile
boundaries, so the cells are of unequal size. $\Delta CS$ is in dollars per
Medicare eligible per month, eligible-weighted within quintile. Substitution
parameters are held at the preferred estimates, $\sigma$ of 0.136 for HMO
nests and 0.117 for PPO nests with $\alpha{=}{-}0.0336$, and the rows vary
the crowding parameter alone, as in Table~\ref{tab:tau_sensitivity}.
The ratio row divides the $\tau{=}0$ mean by the preferred mean within
quintile.
Unchanged menus are markets whose 2024 plan count does not exceed their 2018
count, so the counterfactual removes nothing and $\Delta CS$ is zero by
construction.
\end{minipage}
\end{table}

Scaling the mean
per-eligible monthly figure by the Medicare-eligible population of the sample
counties gives an aggregate of roughly \$518.5 million per year. We report this
as illustrative: it holds prices, unobserved quality, and the nesting
parameters fixed, and it assumes the per-eligible surplus change applies
uniformly within each county.

\FloatBarrier
\section{Conclusion} \label{sec:conclusion}
The exact likelihood of the Berry [1994] nested-logit share inversion includes the Jacobian term $\sum_m\sum_g(J_{gm}-1)\ln(1-\sigma_g)$. Because this term depends on $\sigma_g$, omitting it changes the likelihood estimate even when nest sizes are fixed. When nest sizes vary, the omitted contribution also varies across markets.

The corrected likelihood identifies the nesting parameter without an instrument for the within-nest share. This matters when product counts enter mean utility. In that case, count instruments fail the exclusion restriction, while the corrected likelihood permits a product-count term to be included directly in demand. Implementation requires one additional term in the objective function; price endogeneity remains a separate problem.

In the Medicare Advantage application, count IV produces an inadmissible estimate in the single-nest model and substantially larger substitution parameters than the corrected likelihood in the two-nest model. The estimated product-count term is negative. Accounting for it reduces the estimated 2024 consumer-surplus gain from post-2019 plan growth by 43 percent and implies losses in a small share of high-count markets.

The likelihood correction and the direct product-count term address different parts of the model. The Jacobian corrects estimation of within-nest substitution; the product-count term determines how additional products affect mean utility and welfare. Both are required when evaluating changes in product variety with variable choice sets.

\bibliographystyle{plainnat}
\bibliography{references}

\end{document}